\documentclass[aps,pra,twocolumn]{revtex4-2}
\usepackage{amsmath,amsfonts,amsthm,bbm}
\usepackage[dvipsnames]{xcolor}
\usepackage{graphicx}
\usepackage{multirow,mhchem,chemformula}
\usepackage[colorlinks=true,breaklinks=true,linkcolor=red,citecolor=magenta,urlcolor=magenta]{hyperref}

\graphicspath{ {./Images/} }

\usepackage{tikz}
\usetikzlibrary{quantikz}

\newcommand{\OHradical}{\ch{OH}^\bullet}
\newcommand{\OHanion}{\ch{OH}^-}
\newcommand{\crt}[1]{\hat{c}_{#1}^\dagger}
\newcommand{\dst}[1]{\hat{c}_{#1}^{\phantom{\dagger}}}
\newcommand*\vett[1]{{\bf{#1}}}

\newcommand{\device}[1]{$\mathsf{ibm\_#1}$}
\newcommand*{\citen}[1]{%
  \begingroup
    \romannumeral-`\x 
    \setcitestyle{numbers}%
    \cite{#1}%
  \endgroup   
}

\begin{document}

\title{$N$-electron valence perturbation theory with reference wavefunctions from quantum computing:
application to the relative stability of hydroxide anion and hydroxyl radical}

\author{Alessandro Tammaro}
\affiliation{Universit\`a degli Studi di Milano, Dipartimento di Fisica ``Aldo Pontremoli'', via Celoria 16, I-20133 Milano, Italy}
\author{Davide E. Galli}
\affiliation{Universit\`a degli Studi di Milano, Dipartimento di Fisica ``Aldo Pontremoli'', via Celoria 16, I-20133 Milano, Italy}
\author{Julia E. Rice}
\affiliation{IBM Quantum, IBM Research Almaden, 650 Harry Road, San Jose, CA 95120, USA}
\author{Mario Motta}
\affiliation{IBM Quantum, IBM Research Almaden, 650 Harry Road, San Jose, CA 95120, USA}

\begin{abstract}
Quantum simulations of the hydroxide anion and hydroxyl radical are reported, employing variational quantum algorithms for near-term quantum devices.
The energy of each species is calculated along the dissociation curve, to obtain information about the stability of the molecular species being investigated. 
It is shown that simulations restricted to valence spaces incorrectly predict the hydroxyl radical to be more stable than the hydroxide anion.
Inclusion of dynamical electron correlation from non-valence orbitals is demonstrated,
through the integration of the variational quantum eigensolver and quantum subspace expansion methods in the workflow of $N$-electron valence perturbation theory, 
and shown to correctly predict the hydroxide anion to be more stable than the hydroxyl radical, provided that basis sets with diffuse orbitals are also employed.
Finally, we calculate the electron affinity of the hydroxyl radical using
an aug-cc-pVQZ basis on IBM's quantum devices.
\end{abstract}

\maketitle

\section{Introduction}


The simulation of many-body quantum systems is an important application for a quantum computer \cite{georgescu2014quantum,cao2019quantum,cerezo2020variational,bauer2020quantum,mcardle2020quantum,motta2021emerging}. 
In the context of quantum chemistry, an important example of such an application is the electronic structure problem, 
namely solving for the ground or low-lying eigenstates of the electronic Schr\"{o}dinger equation for the Born-Oppenheimer hamiltonian \cite{friesner2005ab,helgaker2012recent,helgaker2014molecular}.

In recent years, a variety of quantum algorithms has delivered promising results in the calculation of potential energy curves,
ground- and excited-state energies and ground-state correlation functions for a variety of molecules \cite{cao2019quantum,cerezo2020variational,bauer2020quantum,mcardle2020quantum,motta2021emerging}.
Notwithstanding this progress, the limitations of contemporary quantum computation platforms 
have resulted in most quantum electronic structure simulations reported to date employing minimal basis sets (i.e. describing core and valence orbitals only)
or being restricted to active spaces of a few orbitals and electrons.
While these simulations include some electronic correlation, thanks to the ability to entangle electrons within the active space, 
the dynamical correlation arising from inactive orbitals is important to obtain quantitatively and qualitatively correct results.

In recent years, a number of hybrid quantum-classical algorithms have been proposed, 
which aim to combine simulations on contemporary quantum computation platforms with pre- and post-processing operations carried out on classical computers, 
in order to achieve more expressive computations
\cite{bravyi2016trading,kreula2016few,yamazaki2018towards,peng2020simulating,takeshita2020increasing,kawashima2021optimizing,mitarai2021constructing,yuan2021quantum,eddins2021doubling}.

In the present work, we integrate the variational quantum eigensolver (VQE) \cite{farhi2014quantum,peruzzo2014variational,mcclean2016theory,romero2018strategies}
and quantum subspace expansion (QSE) \cite{mcclean2017hybrid,colless2018computation,huggins2020non}
methods in the workflow of N-electron valence perturbation theory (NEVPT2) \cite{angeli2001introduction,angeli2001n,sokolov2016time,sokolov2017time}.

The combination of VQE and QSE gives an approximation for the ground and excited states of the Born-Oppenheimer Hamiltonian within an active space of valence orbitals and electrons
based on intrinsic atomic orbitals (IAOs) \cite{knizia2013intrinsic,senjean2020generalization,schwilkIAO,ManzIAO,WestIAO,ElviraIAO,SchneiderIAO,barison2020quantum}.
Information from these calculations is then used to compute a perturbative correction to the ground-state energy provided by VQE,
that accounts for one- and two-electron transitions from active to inactive orbitals.
We apply the NEVPT2 formalism to examine the relative stability of the hydroxide anion and hydroxyl radical. 

Although the hydroxyl radical is known experimentally to strongly bind an electron
\cite{celotta1974laser,hotop1974high,schulz1982oh}, Hartree-Fock calculations \cite{cade1967hartree} predict the excess electron to be unbound. 
The electron affinity of hydroxyl radical is therefore entirely due to differential effects of electron correlation between the neutral and the anion, a feature that makes the problem
particularly interesting for theoretical calculations \cite{smith1974theoretical,meyer1974pno,sasaki1974configuration,rosmus1978pno,botch1982theoretical,novoa1985electron,raghavachari1985basis,baker1986evaluation,chipman1986electron}.

The remainder of the present work is organized as follows. The NEVPT2 formalism and its integration with VQE and QSE are described in Section \ref{sec:methods}.
Results are presented in Section \ref{sec:results}, conclusions are drawn in Section \ref{sec:conclusions}, and an appendix reports additional computational details.

\section{Methods}
\label{sec:methods}

We begin with a brief overview of multi-reference perturbation theory, and an instructional account of the working equations used in the present study. 
Our starting point is the Born-Oppenheimer Hamiltonian written in second quantization (Chemists notation),
\begin{equation}
\hat{H} = h_0^\prime + \sum_{\substack{p^\prime r^\prime \\ \sigma}} h_{p^\prime r^\prime} \crt{p^\prime \sigma} \dst{r^\prime \sigma} + \sum_{\substack{p^\prime r^\prime q^\prime s^\prime \\ \sigma\tau}} \frac{v_{p^\prime r^\prime q^\prime s^\prime}}{2} \crt{p^\prime \sigma} \crt{q^\prime \tau} \dst{s^\prime\tau}  \dst{r^\prime \sigma}
\end{equation}
where indices {$p^\prime,r^\prime,q^\prime,s^\prime$} label spatial orbitals in a finite orthonormal basis, and $\sigma,\tau \in \{ \uparrow,\downarrow \}$ are spin indices. The nucleus-nucleus Coulomb interaction is described by
\begin{equation}
h_0^\prime = \sum_{\alpha < \beta }^{N_{nuc}} \frac{Z_\alpha Z_\beta}{\| \vett{R}_\alpha - \vett{R}_\beta \|} \quad,
\end{equation}
where $\vett{R}_\alpha$ and $Z_\alpha$ are the position and atomic number of nucleus $\alpha$.
The coefficients 
\begin{equation}
\begin{split}
h_{p^\prime r^\prime} &= \int d {\bf{r}} \, \varphi_{p^\prime} ({\bf{r}}) \, \left[ - \frac{1}{2} \, \frac{\partial^2}{\partial \vett{r}^2}  - \sum_{\alpha=1}^{N_{nuc}} \frac{Z_\alpha }{ \| \vett{r} - \vett{R}_\alpha \| } \right] \, \varphi_{r^\prime}({\bf{r}})
\\
v_{p^\prime r^\prime q^\prime s^\prime} &= \int d {\bf{r}}_1 \int d {\bf{r}}_2 \,  \frac{ \varphi_{p^\prime} ({\bf{r}}_1) \varphi_{r^\prime} ({\bf{r}}_1) \, \varphi_q^\prime ({\bf{r}}_2) \varphi_{s^\prime} ({\bf{r}}_2) }{ r_{12} }
\end{split}
\end{equation}
describe the one-electron part of the Hamiltonian and the electron-electron Coulomb interaction respectively. 
Hartree units are used throughout, the numbers of spin-up and spin-down electrons and nuclei are $N_\uparrow$, $N_\downarrow$, and $N_{nuc}$ respectively,
and orbitals $\varphi_p$ are assumed real-valued, which ensures $(pr|qs)$ has 8-fold symmetry.

Following published literature \cite{sokolov2016time,sokolov2017time}, we partition the spatial orbitals into three sets:
(i) core (doubly-occupied) with indices $i,j,k$ 
(ii) active with indices $t,u,v,w$ and 
(iii) external (unoccupied) with indices $a,b,c$. 
We construct core, active, and external orbitals with a procedure based on the formalism of IAOs \cite{knizia2013intrinsic}.
IAOs are localized molecular orbitals arising 
from a simple algebraic construction, free from input from first-principle numerical simulations, that can be used to define atomic core and valence orbitals, polarized by
the molecular environment. These orbitals can exactly represent self-consistent field wave functions. As IAOs span the molecular valence space, they represent a natural starting point for perturbative inclusion of single and double excitations into external orbitals. See Appendix \ref{app:orbitals} for more details.

In this case the 1s core orbital of oxygen is frozen, leading to the transformed Hamiltonian
\begin{equation}
\label{eq:core_frozen}
\hat{H} = h_0 + \sum_{\substack{pr \\ \sigma}} h_{pr} \crt{p\sigma} \dst{r\sigma} + \sum_{\substack{prqs \\ \sigma\tau}} \frac{v_{prqs}}{2} \crt{p\sigma} \crt{q\tau} \dst{s\tau}  \dst{r\sigma}
\end{equation}

where the coefficients $h_0$, $t_{pr}$ and $v_{prqs}$ are detailed in Appendix \ref{app:core} and the indices $p,r,q,s$ are used to indicate
active or external orbitals. The Hamiltonian is written as the sum of a Dyall operator \cite{dyall1995choice},
\begin{equation}
\label{eq:dyall}
\hat{H}_d = \sum_{\substack{a \\ \sigma}} \varepsilon_a \, \crt{a\sigma} \dst{a\sigma} + \hat{H}_{act} \quad,
\end{equation}
and of a perturbation $\hat{V} = \hat{H} - \hat{H}_d$. 
In Eq.~\eqref{eq:dyall}, the orbital energies $\varepsilon_a$ are defined as the eigenvalues of the projection of the Fock operator $\hat{F}$ on the external space, and
\begin{equation}
\hat{H}_{act} = h_0 + \sum_{\substack{tu \\ \sigma}} h_{tu} \crt{t\sigma} \dst{u\sigma} 
+ \sum_{\substack{tuvw \\\sigma\tau}} \frac{v_{tuvw}}{2} \crt{t\sigma} \crt{u\tau} \dst{v\tau}  \dst{w\sigma}
\end{equation}
is the restriction of the Born-Oppenheimer Hamiltonian to the active space.
The second-order energy contribution can be written as
\begin{equation}
\label{eq:nevpt2}
- E_{\mathrm{PT2}} = \sum_{\nu \neq 0} \frac{| \langle \Psi_\nu | \hat{V} | \Psi_0 \rangle |^2}{E_\nu - E_0}
\end{equation}
where $(\Psi_\nu,E_\nu)$ are the eigenpairs of the Dyall Hamiltonian $\hat{H}_d$, where  $\nu=0$ corresponds to the ground state. 
Eq~\eqref{eq:nevpt2} is the second-order energy expression from the Rayleigh-Schr\"{o}dinger perturbation theory,
which yields the exact energy of the second-order $N$-electron valence perturbation theory (NEVPT2) \cite{angeli2001introduction,angeli2001n,sokolov2016time,sokolov2017time}.

In order to evaluate Eq.~\eqref{eq:nevpt2}, it is necessary to know all the eigenvalues and eigenvectors of the Dyall Hamiltonian 
such that $\langle \Psi_\nu | \hat{V} | \Psi_0 \rangle \neq 0$. To elucidate the structure of such eigenstates, it is useful to recall
that the action of $\hat{V}$ over the ground state reads
\begin{equation}
\label{eq:nevpt3}
\begin{split}
\hat{V} | \Psi_0 \rangle 
&= 
\Bigg[ \sum_{\substack{a \\ \sigma}} \crt{a\sigma} \hat{O}^{(1)}_{a,\sigma} + \sum_{\substack{a<b \\ \sigma}} \crt{a\sigma}\crt{b\sigma} \hat{O}^{(2)}_{ab,\sigma} \\
&+ 
\sum_{ab} \crt{a\uparrow} \crt{b\downarrow} \hat{O}^{(3)}_{ab} \Bigg] | \Psi_0 \rangle \\
\end{split}
\end{equation}
where the operators
\begin{equation}
\label{eq:nevpt4}
\begin{split}
\hat{O}^{(1)}_{a,\sigma} &= \sum_{t} h_{at} \, 
\dst{t\sigma} + \sum_{ \substack{ tuv\\ \tau} } v_{atuv} \,
\crt{u\tau} \dst{v\tau} \dst{t\sigma} \quad, \\
\hat{O}^{(2)}_{ab,\sigma} &= \sum_{tu} v_{atbu} \, \dst{u\sigma}  \dst{t\sigma} \quad, \\
\hat{O}^{(3)}_{ab} &= \sum_{tu} v_{atbu} \, \dst{u\downarrow}  \dst{t\uparrow} \quad, \\
\end{split}
\end{equation}
respectively remove a particle with spin $\sigma$, two particles with identical spins $\sigma$, and two particles with opposite spin from the active space.
In the light of Eq.~\eqref{eq:nevpt4}, Eq.~\eqref{eq:nevpt2} takes the form
\begin{equation}
\begin{split}
\label{eq:nevpt5}
- E_{\mathrm{PT2}}
&= \sum_{\lambda} \sum_{\substack{a \\ \sigma}} 
\;
\frac{| \langle \Phi^{(\sigma)}_\lambda | \hat{O}^{(1)}_{a,\sigma} | \Phi_0 \rangle |^2}{\varepsilon_a + \tilde{E}^\sigma_\lambda - \tilde{E}_0} \\
&+ \sum_{\lambda} \sum_{\substack{a<b \\ \sigma}} 
\;
\frac{| \langle \Phi^{(\sigma\sigma)}_\lambda | \hat{O}^{(2)}_{ab,\sigma} | \Phi_0 \rangle |^2}{\varepsilon_a + \varepsilon_{b} + \tilde{E}^{\sigma\sigma}_\lambda - \tilde{E}_0} \\
&+ \sum_{\lambda} \sum_{ab} 
\;
\frac{| \langle \Phi^{(\uparrow\downarrow)}_\lambda | \hat{O}^{(3)}_{ab} | \Phi_0 \rangle |^2}{\varepsilon_a + \varepsilon_{b} + \tilde{E}^{\uparrow\downarrow}_\lambda - \tilde{E}_0} \\
\end{split}
\end{equation}
where $\tilde{E}_0$ and $\Phi_0$ denote the ground-state energy and wavefunction of $\hat{H}_{act}$, and $\Phi_0$ has $(N_\uparrow,N_\downarrow)$ particles. In addition,
$(\Phi_\lambda^{(\uparrow)},\tilde{E}^{(\uparrow)}_\lambda)$,
$(\Phi_\lambda^{(\uparrow\downarrow)},\tilde{E}^{(\uparrow\downarrow)}_\lambda)$
denote the eigenpairs of the active-space Hamiltonian $\hat{H}_{act}$ 
in the sectors of the Fock space with 
$(N_\uparrow-1,N_\downarrow)$,
$(N_\uparrow-1,N_\downarrow-1)$ particles,
etc.

Unlike Eq.~\eqref{eq:nevpt2}, the last expression for the correlation energy involves solutions of the Schr\"{o}dinger equation in the active space.
A natural way to approximately evaluate Eq.~\eqref{eq:nevpt5} is to integrate the variational quantum eigensolver (VQE) 
and quantum subspace expansion methods (QSE) in the workflow of NEVPT2. More specifically, 

(i) an initial VQE calculation is performed, to approximate the ground state of $\hat{H}_{act}$; (ii) then, the following Ans\"{a}tze are formulated for the excited states,
\begin{equation}
\label{eq:nevpt6}
\begin{split}
| \Phi^\sigma_{\lambda} \rangle &= \sum_u \omega^\sigma_{u,\lambda} \dst{u\sigma} | \Phi_0 \rangle \quad, \\
| \Phi^{\sigma\sigma}_{\lambda} \rangle &= \sum_{u<v} \omega^{\sigma\sigma}_{uv,\lambda} \dst{u\sigma} \dst{v\sigma} | \Phi_0 \rangle \quad, \\
| \Phi^{\uparrow\downarrow}_{\lambda} \rangle &= \sum_{uv} \omega^{\uparrow\downarrow}_{uv,\lambda} \dst{u\uparrow} \dst{v\downarrow} | \Phi_0 \rangle \quad. \\
\end{split}
\end{equation}

(ii) the energies $\tilde{E}^\sigma_\lambda$ and coefficients $\omega^\sigma_{u,\lambda}$ are evaluated by forming the overlap and Hamiltonian matrices 
\begin{equation}
\label{eq:nevpt7}
\begin{split}
S^\sigma_{uv} &= \langle \Phi_0 | \crt{u\sigma} \dst{v\sigma} | \Phi_0 \rangle \quad, \\
H^\sigma_{uv} &= \langle \Phi_0 | \crt{u\sigma} \hat{H}_{act} \dst{v\sigma} | \Phi_0 \rangle \quad, \\
\end{split}
\end{equation}
and solving the eigenvalue equation
\begin{equation}
\label{eq:nevpt7b}
\sum_v H^\sigma_{uv} \omega^\sigma_{v,\lambda} = \tilde{E}^\sigma_\lambda \sum_v S^\sigma_{uv} \omega^\sigma_{v,\lambda} \quad .
\end{equation}
An analogous procedure is carried out to compute the energies
$\tilde{E}^{\sigma\sigma}_\lambda$, $\tilde{E}^{\uparrow\downarrow}_\lambda$
and the coefficients
$\omega^{\sigma\sigma}_{uv,\lambda}$, $\omega^{\uparrow\downarrow}_{uv,\lambda}$.

(iii) the transition matrix elements appearing in Eq.~\eqref{eq:nevpt5} are computed with the formulas reported in Appendix \ref{app:transition}, and $E_{\mathrm{PT2}}$ is evaluated.

\subsection{Computational cost and accuracy limitations}

We now quantify the computational cost of the procedure outlined in the previous Section. We denote with $N_{act}$ the number of active orbitals
and with $N_{qse}$ the number of QSE states, which is $N_{act}$, $(N_{act}^2-N_{act})/2$, $N_{act}^2$ for the three sets of states
in Eq.~\eqref{eq:nevpt6}.

Computing the QSE overlap and Hamiltonian matrices exemplified in  Eq.~\eqref{eq:nevpt7} requires $\mathcal{O}(N_{act}^4 N_{qse}^2)$ measurements of Pauli operators.
Solving the eigenvalue equation~\eqref{eq:nevpt7b} requires $\mathcal{O}(N_{qse}^3)$ flops on a classical computer.

Computing the transition matrix elements in Eq.~\eqref{eq:nevpt5}
require $\mathcal{O}(N^2_{ext} N^2_{act} N_{qse})$ flops on a classical computer, as explained in Appendix \ref{app:transition}. 
Computing $\Delta E$ requires $\mathcal{O}(N_{qse} N_{ext}^2)$ flops on a classical computer.

The overall cost is of $\mathcal{O}(N_{act}^4 N_{qse}^2)=\mathcal{O}(N_{act}^8)$ Pauli measurements on a quantum computer
and, since in general $N_{ext}$ is in general much larger than $N_{act}$,
of $\mathcal{O}(N_{ext}^2)$ additional operations on a classical computer.
The Ansatz 
Eq.~\eqref{eq:nevpt6} 
introduces two approximations with respect to NEVPT2 \cite{angeli2001introduction}: 
first, the replacement of the exact ground state (GS) with a VQE Ansatz; second, the retention of a limited number of excited states (ES).
In the remainder of this work, we endeavor to assess the impact of both approximations on the final results, by comparing:

(i) NEVPT2 with exact GS and exact ES, denoted NEVPT2(FCI,FCI), 

(ii) NEVPT2 with exact GS, and ES approximated by Eq.~\eqref{eq:nevpt6}, denoted NEVPT2(FCI,QSE), 

(iii) NEVPT2 with VQE Ansatz, and ES approximated by Eq.~\eqref{eq:nevpt6}, denoted NEVPT2(Ansatz,QSE).

Comparison of (i) versus (ii), and (ii) versus (iii), provides a way to assess 
the impact of the QSE and VQE approximations on the accuracy of NEVPT2, respectively.

\subsection{Additional computational details}

The calculations performed in this work involved initial pre-processing by the quantum chemistry code PySCF \cite{sun2018pyscf,sun2020recent}) 
on classical computers, to generate optimized mean-field orbitals and Hamiltonian coefficients prior to performing computations with quantum simulators. 
The restricted closed- and open-shell Hartree-Fock (RHF and ROHF respectively, also denoted SCF) states were chosen as the initial states for all of the calculations described here. 
We compared SCF calculations with correlated calculations
employing M${\o}$ller-Plesset perturbation theory (MP2),
coupled-cluster with singles and doubles 
and perturbative triples (CCSD and CCSD(T) respectively),
and full configuration interaction (FCI or exact diagonalization) \cite{helgaker2014molecular}.
All correlated calculations used the frozen core approximation (1s orbital for oxygen).
This leads to $N_{act} = 5$ orbitals for all basis sets, and $N_{ext}$ ranging from 11 (6-31++G) to  121 (aug-cc-pVQZ) orbitals.

Having selected a set of single-electron orbitals for each of the studied species, VQE computations were performed with quantum simulators.
We used IBM's open-source library for quantum computing, Qiskit \cite{aleksandrowicz2019qiskit}. Qiskit 
contains implementations of techniques to map the fermionic Fock space onto the Hilbert space of a register of qubits, and an implementation of the VQE algorithm.
Here we use the tapering-off technique \cite{bravyi2017tapering,setia2019reducing} to account for molecular orbital point group symmetries which  reduces the number of qubits required for a simulation. 
In analogy with conventional symmetry-adapted quantum chemistry calculations, this reduction does not introduce additional approximations in the calculations.
For the systems considered here, the tapering-off technique reduced the number of qubits to $n_q=6$.

In the VQE algorithm, we took our wavefunction in the form of a quantum circuit, which was either the quantum unitary coupled cluster with singles 
and doubles q-UCCSD as implemented in Ref. \cite{barkoutsos2018quantum}, or the following $R_y$ Ansatz,
\begin{equation}
\begin{split}
| \Psi(\theta) \rangle &= 
\prod_{k=1}^{n_r}
\left[
\prod_{i=0}^{n_q-1} R^{i}_{y}(\theta_{k}^i)
E
\right]
\prod_{i=0}^{n_q-1} R^{i}_{y}(\theta_{0}^i) | \Psi_{init} \rangle
\;, \\
E &= \prod_{i=0}^{n_q-2} c_i X_{i+1} \;,
\end{split}
\end{equation}
where $| \Psi_{init} \rangle$ is an initial wavefunction (here, the restricted closed- or open-shell Hartree-Fock state), $n_q$ is the number of qubits, $R^i_{y}(\theta) 
= \mbox{exp}(-i \theta Y_i/2)$ is a $Y$ rotation of an angle $\theta$ applied to qubit $i$, $c_i X_{i+1}$ is a CNOT gate with control and target qubits, $i$ and $i+1$ respectively,
and $n_r$ is an integer denoting the number of times a layer of entangling gates followed by a layer of $Y$ rotations is repeated. 
In this study, to ensure an accurate representation of the ground-state wavefunction by the $R_y$ Ansatz, we chose $n_r = 3$, corresponding to the quantum circuit shown in Appendix \ref{app:ry}.

We then minimized the expectation value of the Hamiltonian with respect to the parameters,  $\theta$ in the circuit. 
The minimization was carried out using the classical optimization method, L-BFGS-B \cite{zhu1997algorithm,morales2011remark}. 
We ran our experiments on the ideal statevector simulator of Qiskit.
Once the VQE had completed, we obtained the optimized variational form and the estimate for the ground state energy.
In addition, we measured the operators required to construct the QSE overlap and Hamiltonian matrices, Eq.~\eqref{eq:nevpt7}.

We performed hardware experiments on IBM's 27-qubit processors \device{kolkata} and
\device{auckland} based on the Falcon architecture.
We employed readout error mitigation \cite{temme2017error,kandala2019error,bravyi2020mitigating} as implemented in Qiskit 
Runtime \cite{nation2021scalable} to correct measurement errors. 
We also used a zero-noise extrapolation method introducing additional CNOT gates to 
account for errors introduced during the expensive 2-qubit entangling operations,
as described in Refs~\citen{dumitrescu2018cloud,stamatopoulos2019option}.


\begin{figure*}[t!]
\includegraphics[width=0.87\textwidth]{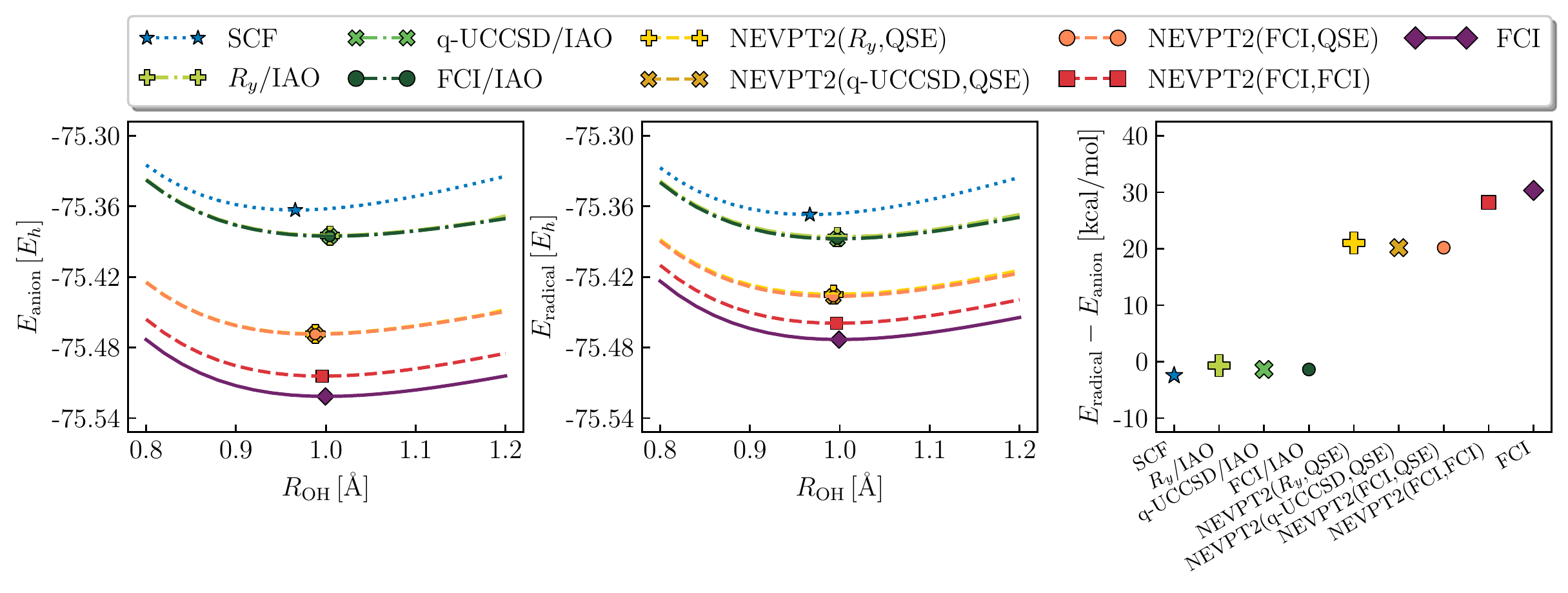}
\caption{Left: potential energy curve of $\OHanion$ (anion)
from SCF (blue dotted line),
VQE and FCI in the IAO basis computed from an underlying 6-31++G basis (green dash-dotted lines),
and various approximations of NEVPT2 (warm colored dashed lines), and FCI in the underlying 6-31++G basis (purple solid line).
Symbols denote equilibrium geometries and energies from a fit of 21 points to a Morse potential.
Middle: same as left, for $\OHradical$ (radical).
Right: Ground-state energy difference between anion and radical.}
\label{figure:631ppg}
\end{figure*}

\begin{figure*}[t!]
\includegraphics[width=0.87\textwidth]{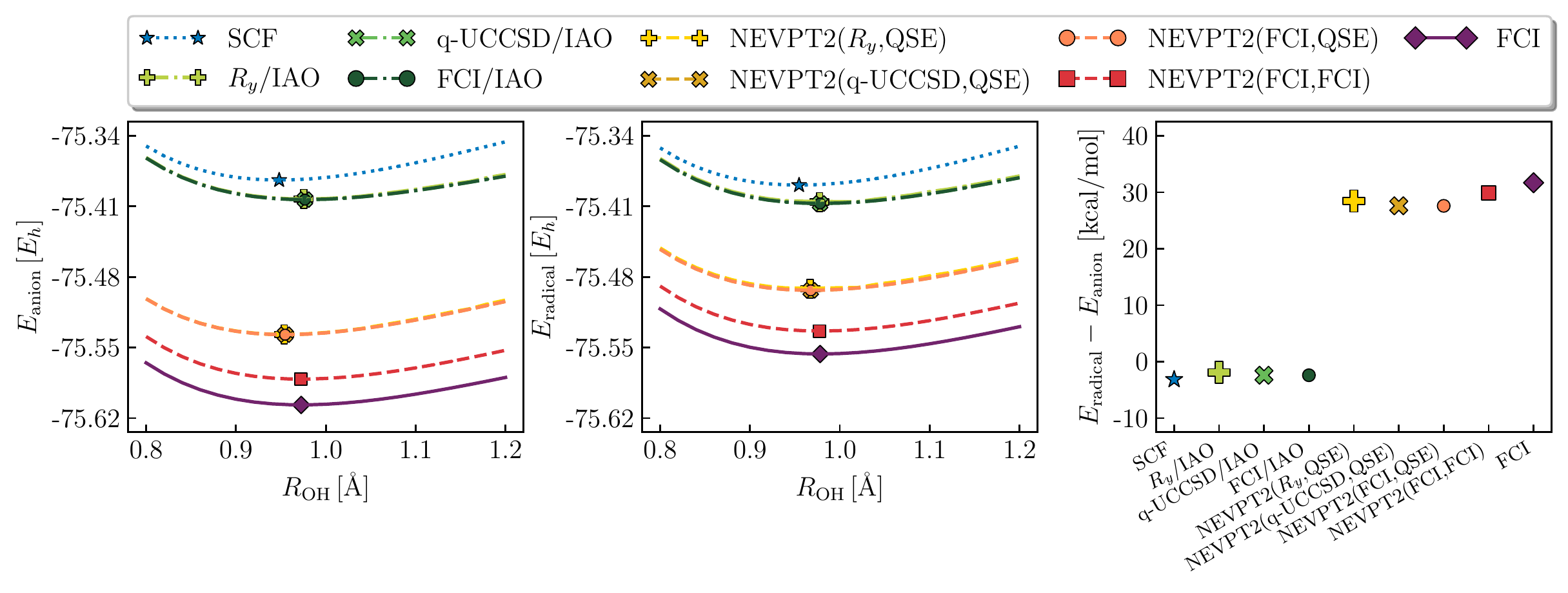}
\caption{Left: potential energy curve of $\OHanion$ (anion)
from SCF (blue dotted line),
VQE and FCI in the IAO basis computed from an underlying 6-31++G${}^{**}$ basis (green dash-dotted lines),
and various approximations of NEVPT2 (warm colored dashed lines), and FCI in the underlying 6-31++G${}^{**}$ basis (purple solid line).
Symbols denote equilibrium geometries and energies from a fit of 21 points to a Morse potential.
Middle: same as left, for $\OHradical$ (radical).
Right: Ground-state energy difference between anion and radical.}
\label{figure:631ppgss}
\end{figure*}


\section{Results}
\label{sec:results}

In this section, we simulated the hydroxide anion ($\OHanion$) and hydroxyl radical ($\OHradical$) using split-valence 6-31G++, and 6-31$++$G${}^{**}$ basis sets
\cite{hehre1972self} and correlation-consistent aug-cc-pVxZ basis sets \cite{dunning1989gaussian}.
For each species, we computed the potential energy curve, namely the ground-state energy as a function of the $\ce{OH}$ bondlength $R_{\mathrm{OH}}$.
We focused on the interval $0.8 \, \mathrm{\AA} \leq R_{\mathrm{OH}} \leq 1.2 \, \mathrm{\AA}$ 
since it includes the experimental gas-phase equilibrium bondlengths of $\OHanion(0.964\mathrm{\AA}$) and $\OHradical (0.970 \, \mathrm{\AA}$) \cite{johnson1999nist}.
For each species, we fit the computed potential energy curve to the Morse potential functional form, 
and extract the equilibrium bondlength $R_{eq} = \mathrm{argmin}_R E(R)$ and the equilibrium ground-state energy $E_{min} = E(R_{eq})$.
We used this information to compute the ground-state energy difference between radical and anion, 
\begin{equation}
\Delta E = E_{\mathrm{radical}}(R_{eq,\mathrm{radical}}) - E_{\mathrm{anion}}(R_{eq,\mathrm{anion}})
\quad.
\end{equation}
which was compared with the experimental electron affinity for the hydroxyl radical of $1.828 \mathrm{eV} (42.1547 \, \mathrm{kcal/mol})$
\cite{johnson1999nist}.


\begin{table*}[t!]
\begin{tabular}{lccc}
\hline\hline
method & $R_{\mathrm{anion}} [\mathrm{\AA}]$ & $R_{\mathrm{radical}} [\mathrm{\AA}]$ & $\Delta E$ [kcal/mol] \\ \hline
SCF & 0.96596(16) & 0.96661(22) & -2.450(23) \\ \hline
$R_y$/IAO & 1.00462(105) & 0.99734(53) & -0.702(86) \\
q-UCCSD/IAO & 1.00442(17) & 0.99720(14) & -1.386(16) \\
FCI/IAO & 1.00442(17) & 0.99720(14) & -1.389(16) \\ \hline
NEVPT2($R_y$,QSE) & 0.98837(81) & 0.99306(57) & 21.030(73) \\
NEVPT2(q-UCCSD,QSE) & 0.98813(13) & 0.99275(13) & 20.187(14) \\
NEVPT2(FCI,QSE) & 0.98813(13) & 0.99274(13) & 20.181(14) \\
NEVPT2(FCI,FCI) & 0.99569(13) & 0.99638(16) & 28.209(16) \\ \hline
FCI & 0.99957(15) & 0.99905(16) & 30.324(16) \\
\hline\hline
\end{tabular}
\caption{Equilibrium bondlengths for $\OHanion$ (anion) and  $\OHradical$ (radical), and energy difference between anion and radical, using the methods defined in Section \ref{sec:methods} and an underlying 6-31++G basis.
Values of $R_{\mathrm{anion}}$ and $R_{\mathrm{radical}}$ reflect the location of symbols in Fig~\ref{figure:631ppg}, left and middle panels. 
Values of $\Delta E$ correspond to the values shown in the right panel of Fig~\ref{figure:631ppg}. The change in the sign of $\Delta E$ indicates the anion is predicted to be more stable than the radical when the full basis is used in NEVPT2 or FCI simulations.}
\label{tab:631ppg}
\end{table*}

\begin{table*}[t!]
\begin{tabular}{lccc}
\hline\hline
method & $R_{\mathrm{anion}} [\mathrm{\AA}]$ & $R_{\mathrm{radical}} [\mathrm{\AA}]$ & $\Delta E$ [kcal/mol] \\ \hline
SCF & 0.94806(9) & 0.95462(19) & -3.161(20) \\ \hline
$R_y$/IAO & 0.97534(56) & 0.97736(84) & -1.919(79) \\
q-UCCSD/IAO & 0.97676(9) & 0.97795(14) & -2.406(13) \\
FCI/IAO & 0.97676(9) & 0.97795(14) & -2.409(13) \\ \hline 
NEVPT2($R_y$,QSE) & 0.95366(46) & 0.96708(89) & 28.445(85) \\
NEVPT2(q-UCCSD,QSE) & 0.95469(6) & 0.96758(12) & 27.602(12) \\
NEVPT2(FCI,QSE) & 0.95469(6) & 0.96757(12) & 27.591(12) \\
NEVPT2(FCI,FCI) & 0.97209(11) & 0.97738(16) & 29.914(16) \\ \hline 
FCI & 0.97255(10) & 0.97806(13) & 31.684(13) \\
\hline\hline
\end{tabular}
\caption{Equilibrium bondlengths for $\OHanion$ (anion) and  $\OHradical$ (radical), and energy difference between anion and radical, using the methods defined in Section \ref{sec:methods} and an underlying 6-31++G$^{**}$ basis.
Values of $R_{\mathrm{anion}}$ and $R_{\mathrm{radical}}$ reflect the location of symbols in Fig~\ref{figure:631ppgss}, left and middle panels. 
Values of $\Delta E$ correspond to the values shown in the right panel of Fig~\ref{figure:631ppgss}.}
\label{tab:631ppgss}
\end{table*}

\subsection{Split-valence bases}

In Fig.~\ref{figure:631ppg} and \ref{figure:631ppgss} we compute the potential energy curve of $\OHanion$ and $\OHradical$
using the split-valence 6-31++G, and 6-31++G${}^{**}$ basis sets respectively.
Numerical values are listed in Tables \ref{tab:631ppg} and \ref{tab:631ppgss}.

As seen, Hartree-Fock incorrectly predicts the radical to be more stable than the anion in all these basis sets, meaning that $\Delta E_{\mathrm{SCF}} < 0$. Both
VQE and FCI simulations carried out in an active space constructed using IAOs increase $\Delta E$, but preserve the incorrect ordering predicted by Hartree-Fock. This is because the diffuse nature of the atomic orbitals in the underlying basis set is mainly reflected in the external orbitals, rather than in the core and valence (active) ones.
Therefore, NEVPT2(FCI,FCI), NEVPT2(FCI,QSE) and NEVPT2(Ansatz,QSE) with $R_y$ or q-UCCSD Ansatz correctly identify the anion as the more stable species, since the underlying basis set contains diffuse functions.

We emphasize that NEVPT2(Ansatz,QSE) are in good agreement with NEVPT2(FCI,QSE) for this simple problem, 
and that the main source of deviations between NEVPT2(FCI,FCI) and NEVPT2(Ansatz,QSE) is the approximation Eq.~\eqref{eq:nevpt6} for excited states.
For the system considered here, the approximation Eq.~\eqref{eq:nevpt6} results in deviations of 4-5 kcal/mol from NEVPT2(FCI,FCI).
On the other hand, NEVPT2(FCI,FCI) results are only 1-2 kcal/mol away from FCI results. A similar trend is seen for equilibrium bondlengths, which are a few m$\mathrm{\AA}$ from FCI results for all basis sets.

Addition of polarization functions on top of diffuse functions from 6-31++G to 6-31++G$^{**}$ improves the agreement between NEVPT2(Ansatz,QSE) and experimental results. 
These quantities differ by 12 kcal/mol when the 6-31++G${}^{**}$ basis is used.
However, such a deviation is naturally expected, given the incompleteness of split-valence bases and the approximations affecting NEVPT2(Ansatz,QSE).


\begin{figure*}[t!]
\includegraphics[width=0.87\textwidth]{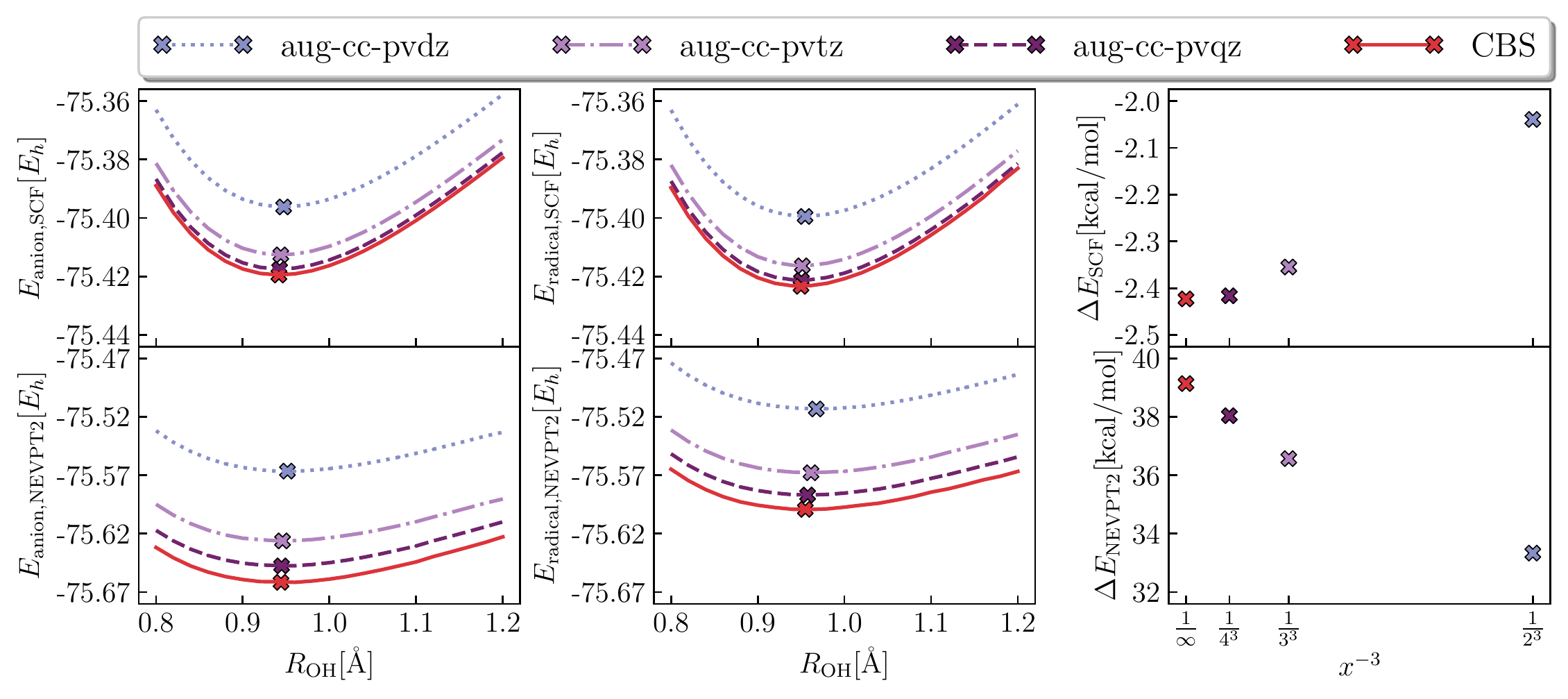}
\caption{Potential energy curves of $\OHanion$ (anion, left) and  $\OHradical$ (radical, middle) from Hartree-Fock (top) and NEVPT2(VQE,QSE)
using Dunning's aug-cc-pVxZ bases with x=D,T,Q (dotted, dot-dashed, and dashed lines respectively) and an $R_y$ Ansatz with linear connectivity for VQE calculations. 
Energies are extrapolated to the complete basis set (CBS, solid red lines) with standard procedures, and crosses denote equilibrium bondlengths and energies.
Right: energy difference between anion and radical from Hartree-Fock (top) and NEVPT2(VQE,QSE) (bottom) as a function of basis set cardinality number $x^{-3}$.
}
\label{figure:dunning}
\end{figure*}

\begin{table*}[t!]
\begin{tabular}{lccc}
\hline\hline
method & $R_{\mathrm{anion}} [\mathrm{\AA}]$ & $R_{\mathrm{radical}} [\mathrm{\AA}]$ & $\Delta E$ [kcal/mol] \\ \hline
SCF       & 0.94174(9) & 0.94994(34) & -2.423(35) \\
\hline
MP2 & 0.96349(22) & 0.96453(22) & 52.733(26) \\
CCSD & 0.95951(11) & 0.96638(20) & 37.826(19) \\
CCSD(T) & 0.96354(11) & 0.96895(15) & 42.294(15) \\
\hline
NEVPT2(R$_y$,QSE) & 0.94413(46) & 0.95488(59) & 39.135(67) \\
NEVPT2(q-UCCSD,QSE) & 0.94211(4) & 0.95591(46) & 38.329(39) \\
NEVPT2(FCI,QSE) & 0.94211(4) & 0.95590(46) & 38.316(39) \\
NEVPT2(FCI,FCI) & 0.96216(10) & 0.96601(23) & 39.446(21) \\
\hline\hline
\end{tabular}
\caption{Equilibrium bondlengths for $\OHanion$ (anion) and  $\OHradical$ (radical), and energy difference between anion and radical, using various classical (SCF, MP2, CCSD, NEVPT2(FCI,QSE), NEVPT2(FCI,FCI), and CCSD(T)) and quantum computing methods (NEVPT2(R$_y$,QSE), NEVPT2(q-UCCSD,QSE)).
Results are extrapolated to the complete basis set limit (CBS) as described in the main text.
The experimental value of $42.1547 \, \mathrm{kcal/mol}$ is taken from Ref.~\cite{johnson1999nist}.}
\label{tab:cbs}
\end{table*}

\subsection{Correlation-consistent augmented bases}

To address the basis set incompleteness error, in Fig.~\ref{figure:dunning} we performed simulations with correlation-consistent augmented bases aug-cc-pVxZ, $x=$D,T,Q or equivalently $2,3,4$.
Hartree-Fock and correlation energies are fit to the exponential Ansatz $E^{\mathrm{HF}}_x = a + b \, e^{-cx}$ with $x=2,3,4$ and the power-law Ansatz $E^{\mathrm{c}}_x = a^\prime + b^\prime \, x^{-3}$ with
$x=3,4$ respectively. This standard procedure  extrapolates the energy to the complete basis set (CBS) limit as $E_{\mathrm{CBS}} = a+a^\prime$ \cite{feller1992application,helgaker1997basis}.

Equilibrium bondlengths and electron affinities extrapolated at CBS level of theory are reported in Table~\ref{tab:cbs}.
As seen, extrapolated equilibrium bondlengths from NEVPT2(FCI,FCI) are within 0.1 Angstrom from both CCSD and experimental values, 
and the QSE approximation introduces additional deviations, of order 0.01 $\mathrm{\AA}$.
Electron affinities from NEVPT2(FCI,FCI) and CCSD are within 1.2 kcal/mol from each other, and 3-4 kcal/mol away from the experimental value.
The QSE approximation causes an additional deviation of 5 kcal/mol from NEVPT2(FCI,FCI) results, which underestimates the electron affinity.

\subsection{Calculations on quantum devices}

\begin{figure*}[t!]
\includegraphics[width=0.9\textwidth]{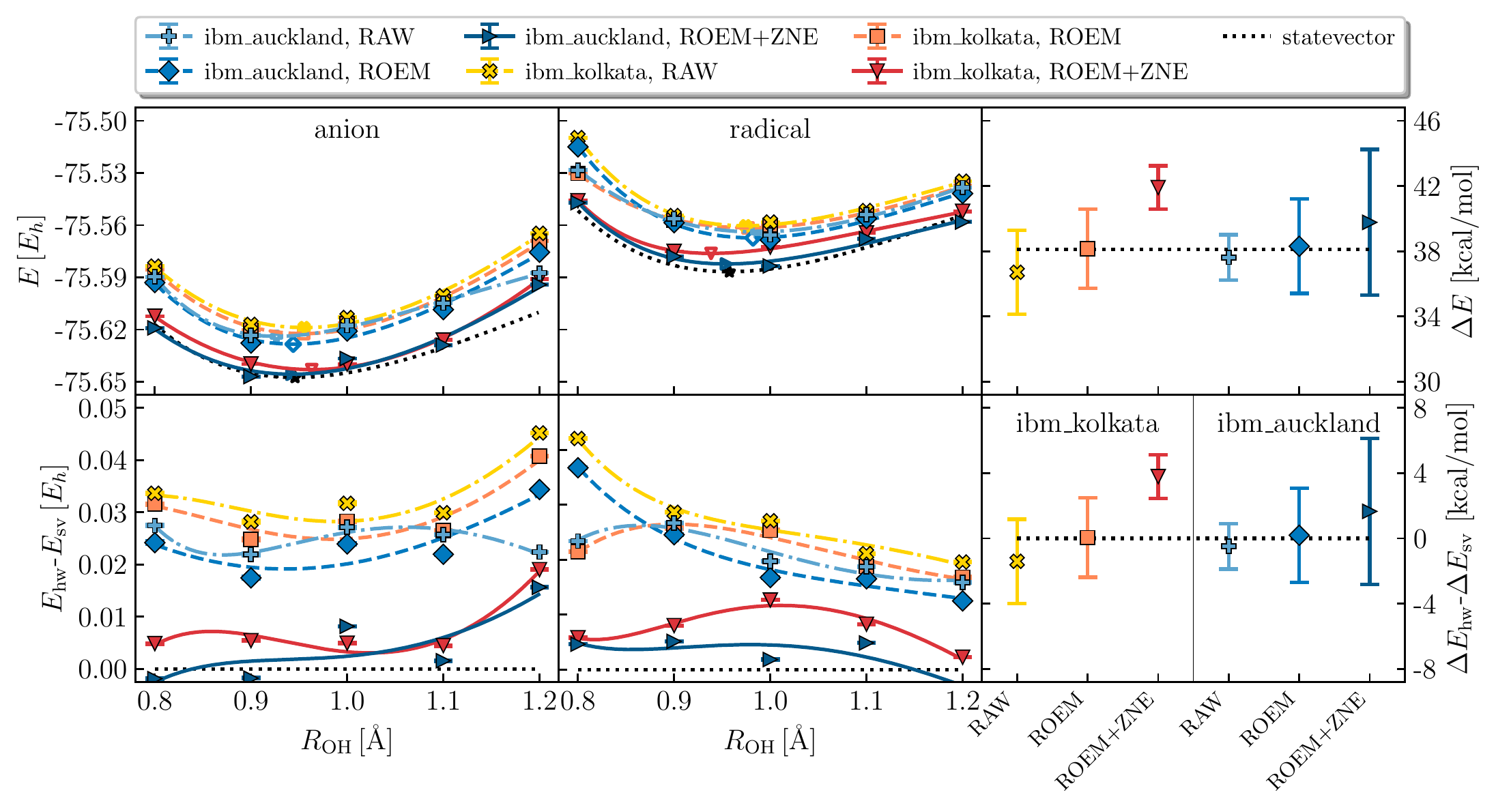}
\caption{
Top: NEVPT2(VQE,QSE) potential energy curves of $\OHanion$ (anion, left) and $\OHradical$ (radical, middle) and electron affinity of $\OHradical$ (right) from noiseless classical simulations (black dotted lines, marked statevector) and quantum hardware \device{auckland} and \device{auckland} (shades of bue and orange respectively), using Dunning's aug-cc-pVQZ basis.
Data without error mitigation (RAW), with readout error mitigated (ROEM), and with ROEM and zero-noise extrapolation (ZNE) are marked by plus, diamond, right-pointing triangle markers on \device{auckland} and cross, square, and 
bottom-pointing triangle markers on \device{auckland}.
Lines denote fit to a Morse potential.
Bottom: Differences between hardware and statevector potential energy curves 
(left, middle) and electron affinity of $\OHradical$ (right).
}
\label{figure:hardware}
\end{figure*}

\begin{figure*}[t!]
\includegraphics[width=0.9\textwidth]{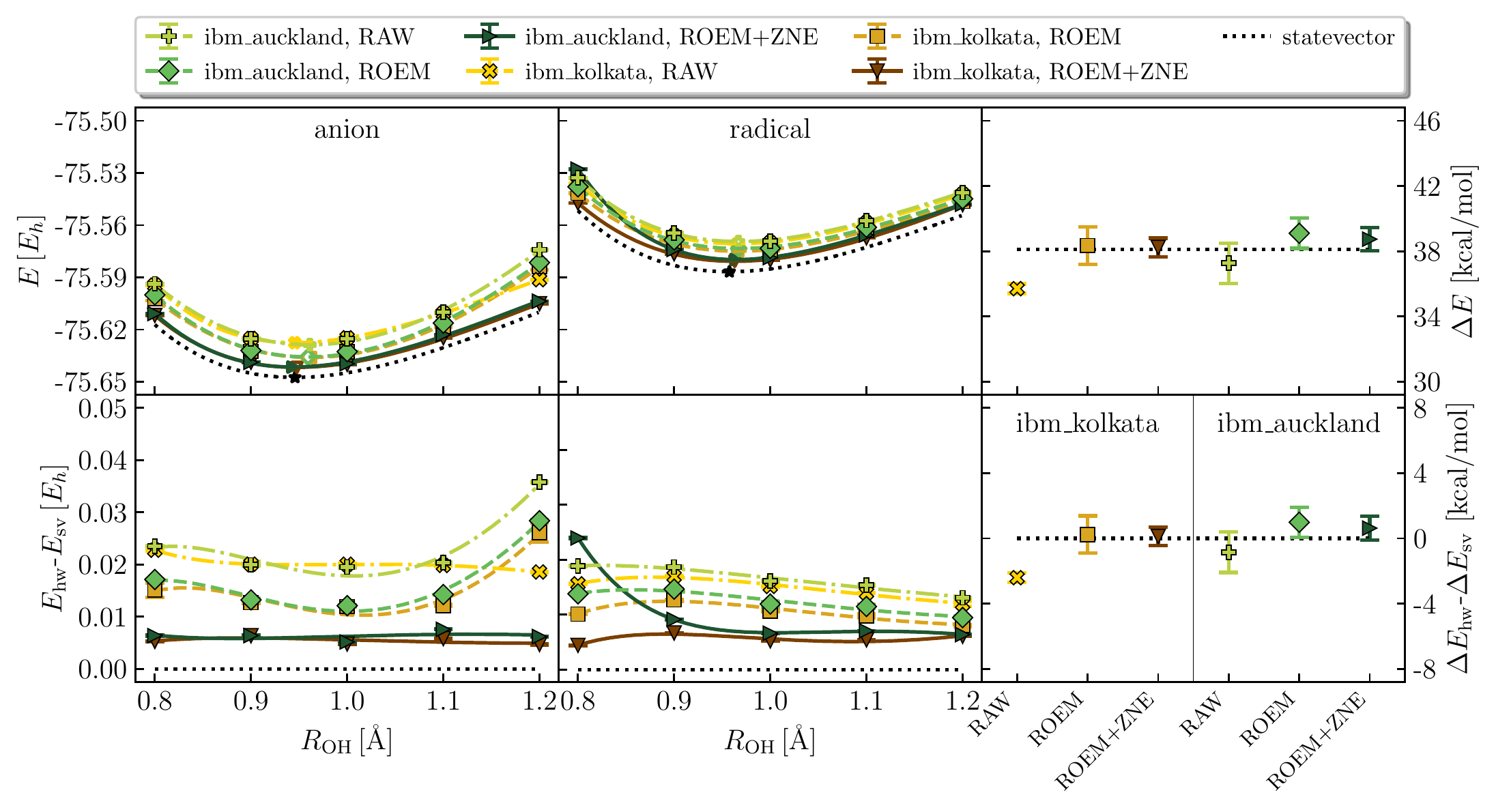}
\caption{
Top: NEVPT2(VQE,QSE) potential energy curves of $\OHanion$ (anion, left) and $\OHradical$ (radical, middle) and electron affinity of $\OHradical$ (right) from classical simulations, both noiseless (black dotted lines, marked statevector) and with noise model from devices \device{auckland} and \device{auckland} (shades of green and brown respectively), using Dunning's aug-cc-pVQZ basis.
Data without error mitigation (RAW), with readout error mitigated (ROEM), and with ROEM and zero-noise extrapolation (ZNE) are marked by plus, diamond, right-pointing triangle markers on \device{auckland} and cross, square, and 
bottom-pointing triangle markers on \device{auckland}.
Lines denote fit to a Morse potential.
Bottom: Differences between hardware and statevector potential energy curves 
(left, middle) and electron affinity of $\OHradical$ (right).
}
\label{figure:qasm}
\end{figure*}

Finally, we evaluate the electron affinity of $\OHradical$ using IBM quantum hardware.
We use an aug-cc-pVQZ basis, we study the five representative bondlengths $R_{\mathrm{OH}}=0.8,0.9,1.0,1.1,1.2$ $\mathrm{\AA}$, and we carry out simulations on IBM's processors \device{kolkata} and \device{auckland} based on the Falcon architecture, as shown in Figure \ref{figure:hardware}.

Results from hardware experiments with and without (ROEM, RAW) readout error mitigation (ROEM) are shown. 
A zero-noise extrapolation (ZNE) \cite{dumitrescu2018cloud,stamatopoulos2019option} is also conducted, with the aim of further mitigating the impact of noise.

As seen in Figure \ref{figure:hardware}, NEVPT2 energies computed on both devices are
$\sim$$300$ milliHartree above statevector results. Deviations between computed and  statevector energies decrease to $\sim$$75$ milliHartree when ROEM and ZNE are used in conjunction.
Nevertheless, the qualitative behavior of both anion and radical is correctly captured
by the hardware experiments upon extrapolation.

We estimate the equilibrium geometries and energies of radical and anion by fitting the
computed energies to a Morse potential, and we estimate the electron affinity of the radical as the difference between such equilibrium energies.
Due to a cancellation of errors, electron affinities are in qualitative agreement with the statevector value, though accompanied by error bars of several kcal/mol.

In Figure \ref{figure:qasm}, we perform simulations analogous to those of Fig.~\ref{figure:hardware}, using a classical simulator (specifically IBM's qasm simulator) with a noise model derived from the calibration of the \device{kolkata} and \device{auckland} processors.
As seen, deviations between simulated and statevector results are less pronounced than in the case of hardware simulations, leading to considerably lower statistical uncertainties 
on fitted quantities. It is understandable that the noise-simulated backends do not faithfully emulate the true hardware noises in our experiments, as the simulated noise models are meant to capture only simple noise channels such a depolarization, amplitude damping, and bit flipping.

\section{Conclusion}
\label{sec:conclusions}

In this work, we integrated the VQE and QSE techniques in the workflow of the NEVPT2 method, and demonstrated such an inclusion
focusing on the relative stability of the hydroxide anion and hydroxyl radical. 
NEVPT2 allows for perturbative inclusion of dynamical correlation arising from non-valence orbitals, 
thereby improving the potential energy curves produced by quantum computing simulations limited to valence spaces. Indeed, simulations in valence spaces by construction capture electronic correlation only within the active space. 
Therefore, perturbative or full inclusion of virtual orbitals is necessary to cover the dynamical correlations with methods like coupled cluster and multireference configuration interaction model, and very important to obtain quantitative agreement with experimental values, especially for sensitive quantities such as polarizabilities or thermochemical properties.

The main limitation of the approach proposed here is the scaling with active space size: it should not be forgotten that the computation of the QSE matrices scales as $\mathcal{O}(N_{act}^8)$, and their diagonalization as $\mathcal{O}(N_{act}^{12})$.

On the other hand, the approach proposed here scales only as $\mathcal{O}(N_{ext}^2)$, due to the perturbative nature of the treatment of external orbitals. Furthermore, it does not involve the additional cost of variationally optimizing the orbitals, as in other approaches \cite{andersson1990second,andersson1992second}.
Therefore, this procedure can capture dynamical correlation energy at reasonable cost with respect to the size of the external space. 

This approach is an example of a hybrid quantum-classical approach using quantum and classical computers in synergy 
to achieve a more accurate result.

We expect that the perturbative inclusion of dynamical correlation from external orbitals, 
as a technique to partially overcome the limitations of calculations employing small basis sets and/or small active spaces, will prove useful in the simulation of chemical species by quantum algorithms on contemporary quantum devices.

\section*{Acknowledgments}

AT, DEG and MM acknowledge the Universit\`a degli Studi di Milano INDACO Platform and the IBM Research Cognitive Computing Cluster service respectively, 
for providing resources that have contributed to the results reported within this study.

\appendix

\section{Computational details}

In this Appendix, we provide additional details about the computational methods used in the present work.

\subsection{Orbital construction}
\label{app:orbitals}

In this Subsection, we describe the construction of core, active, and external orbitals. 
\begin{enumerate}
\item 
First, we choose an underlying basis of atomic orbitals (AOs), 
$\{ \chi_\mu \}_{\mu=1}^M$. 
\item Then, we perform a restricted Hartree-Fock calculation, 
yielding a set of molecular orbitals (MOs),
\begin{equation}
| \psi_k \rangle 
= 
\sum_\mu C^{\mathrm{MO}}_{\mu k} | \chi_\mu \rangle
\quad,\quad k = 1 \dots M
\end{equation}
and a Fock operator $\hat{F}$
\item From the MOs, we construct a set of intrinsic atomic orbitals (IAOs) using standard procedures \cite{sun2018pyscf,sun2020recent},
\begin{equation}
| \eta_f \rangle 
= 
\sum_\mu C^{\mathrm{IAO}}_{\mu f} | \chi_\mu \rangle
\quad,\quad f = 1 \dots N_{iao}
\end{equation}
\item The occupied MOs are by construction \cite{knizia2013intrinsic}
spanned by the IAOs, and can thus be written as
\begin{equation}
\begin{split}
| \psi_i \rangle 
&= 
\sum_f C^{\mathrm{occ}}_{f i} | \eta_f \rangle
\quad,\quad i = 1 \dots N_{\uparrow} \quad, \\
C^{\mathrm{occ}} 
&= \left( S^{\mathrm{IAO}} \right)^{-1} \left[
\left( C^{\mathrm{IAO}} \right)^T S^{\mathrm{AO}} C^{\mathrm{MO}} \right] \quad,
\end{split}
\end{equation}
with $S^{\mathrm{AO}}_{\mu\nu} = \langle \chi_\mu | \chi_\nu \rangle$
and $S^{\mathrm{IAO}}_{fg} = \langle \eta_f | \eta_g \rangle$.

The valence virtual orbitals, which are the orthogonal complement 
of the occupied MOs in the subspace spanned by IAOs, are computed
with a standard Gram-Schmidt procedure,
\begin{equation}
| \xi_l \rangle 
= 
\sum_f C^{\mathrm{vrt}}_{f l} | \eta_f \rangle
\quad,\quad l = N_{iao}-N_{\uparrow} \quad,
\end{equation}
with $\langle \xi_l | \psi_i \rangle = 0$.
\item Core orbitals are the lowest-energy occupied MOs,
\begin{equation}
\label{eq:orb1}
| \varphi_i \rangle = | \psi_i \rangle 
\quad,\quad i=1 \dots N_f \quad.
\end{equation}
Valence orbitals are the non-core occupied MOs and valence virtuals,
\begin{equation}
\label{eq:orb2}
| \varphi_p \rangle = 
\begin{cases}
| \psi_i \rangle \quad,\quad i = N_f+1 \dots N_\uparrow \\
| \xi_l \rangle \quad,\quad l = N_{iao}-N_{\uparrow} \\
\end{cases}
\end{equation}
\item To construct external orbitals, we form the projector 
\begin{equation}
\hat{P}
= \sum_i | \varphi_i \rangle \langle \varphi_i | 
+ \sum_p | \varphi_p \rangle \langle \varphi_p | 
\end{equation}
on the subspace spanned by core and active orbitals, 
and the projector $\hat{Q} = \mathbbm{1} - \hat{P}$
onto its orthogonal complement. We then project the 
Fock operator onto the orthogonal complement of the 
core+active space,
\begin{equation}
\hat{F}^\prime = \hat{Q} \hat{F} \hat{Q} 
\quad.
\end{equation}
External orbitals are the eigenvector of $\hat{F}^\prime$ 
in the orthogonal complement of the core+active space,
\begin{equation}
\label{eq:orb3}
\begin{split}
\hat{F}^\prime   | \varphi_a \rangle 
&= \varepsilon_a | \varphi_a \rangle \quad,\quad 
\hat{Q}          | \varphi_a \rangle 
=               | \varphi_a \rangle \quad. \\
\end{split}
\end{equation}
\end{enumerate}
Equations \eqref{eq:orb1}, \eqref{eq:orb2}, and \eqref{eq:orb3} correspond to the core, active, and external orbitals respectively.

\subsection{Core freezing}
\label{app:core}

In this Subsection we report, for completeness, the standard frozen-core procedure used to remove core orbitals from the simulation.
With the indices $i,j$ and $p,r,q,s$ we respectively denote core and non-core (active or external) orbitals.
\begin{equation}
\begin{split}
\hat{H} &= h_0 + \sum_i 2 h_{ii} + \sum_{\substack{ pr \\ \sigma}} h_{pr} \crt{p\sigma} \dst{r\sigma} \\
&+ \sum_{ij} 2 v_{ii jj} - v_{ ij ji } 
+ \sum_{\substack{ pr \\ \sigma}} \Big[ v_{pr ii} - v_{ir pi} \Big] \crt{p\sigma} \dst{r\sigma} \\
&+ \sum_{\substack{ prqs \\ \sigma\tau}} \frac{v_{pr qs}}{2} \crt{p\sigma} \crt{q\tau} \dst{s\tau}  \dst{r\sigma} \;.
\end{split}
\end{equation}
The Hamiltonian can then be written as in Eq.~\eqref{eq:core_frozen} with
\begin{equation}
\begin{split}
h_0 &= h_0^\prime + \sum_i 2 h_{ii} + \sum_{ij} 2 v_{ii jj} - v_{ ij ji } \quad, \\
h_{pr}^\prime &= h_{pr} + \sum_i v_{pr ii} - v_{ir pi} \quad. \\
\end{split}
\end{equation}

\begin{figure*}[t!]
\includegraphics[width=0.7\textwidth]{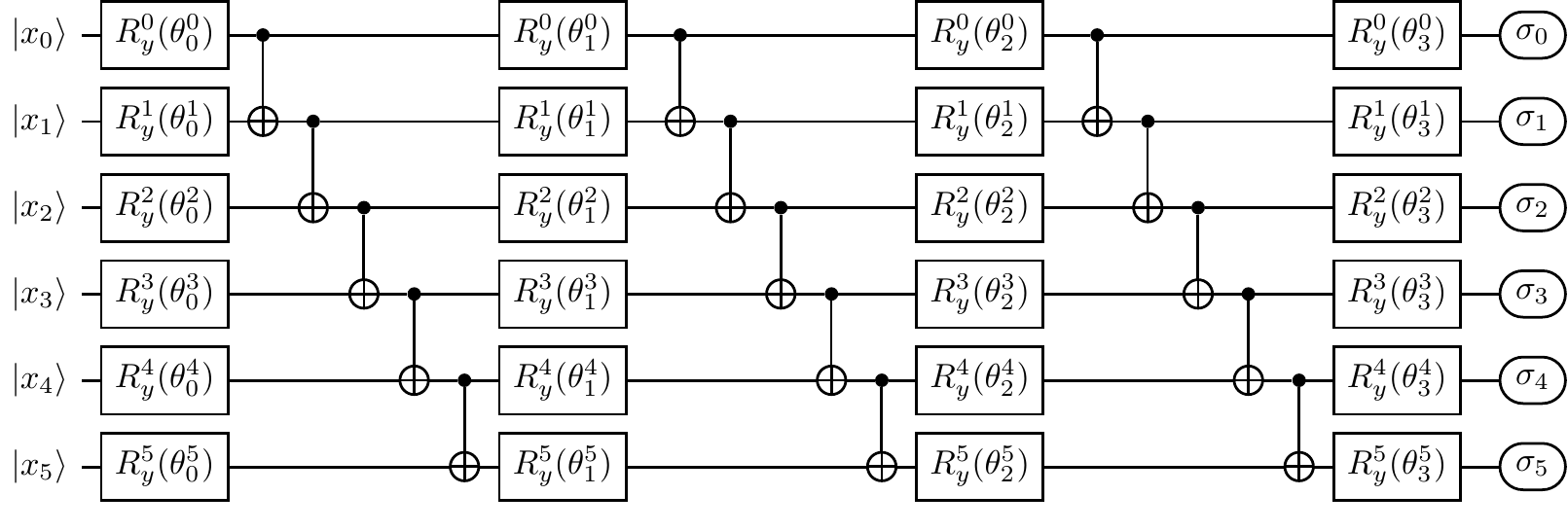}
\caption{Quantum circuit describing the $R_y$ Ansatz with depth $n_r=3$ acting on $n_q=6$ qubits.}
\label{fig:ry}
\end{figure*}

\subsection{Computation of transition matrix elements}
\label{app:transition}

In this Subsection, we detail the computation of the transition matrix elements in Eq.~\eqref{eq:nevpt5}. 
In general, to achieve this goal one has to measure additional operators on a quantum computer. When Eq.~\eqref{eq:nevpt6} is adopted, on the other hand, the outcomes of these additional measurements are trivially related to the QSE overlap matrices.

To verify this point, let us first observe that one- and two-body density matrices are trivially related to the QSE overlap matrices. Indeed,
\begin{equation}
\begin{split}
\rho^\sigma_{uv} 
&= \langle \Phi_0 | \crt{u\sigma} \dst{v\sigma} | \Phi_0 \rangle = S^\sigma_{uv} \quad, \\
\end{split}
\end{equation}
and similarly
\begin{equation}
\begin{split}
\rho^{\sigma\sigma}_{uvwt} 
&= \langle \Phi_0 | \crt{u\sigma} \crt{w\sigma} \dst{t\sigma} \dst{v\sigma} | \Phi_0 \rangle \\
&= (-1)^{\delta_{w w^\prime}+\delta_{t t^\prime}} \, \langle \Phi_0 | \crt{u^\prime\sigma} \crt{w^\prime\sigma} \dst{t^\prime\sigma} \dst{v^\prime\sigma} | \Phi_0 \rangle \\
&= (-1)^{\delta_{w w^\prime}+\delta_{t t^\prime}} S^{\sigma\sigma}_{(w^\prime u^\prime),(t^\prime v^\prime)} \quad, \\
\end{split}
\end{equation}

and

\begin{equation}
\begin{split}
\rho^{\uparrow\downarrow}_{uywz}  
&= \langle \Phi_0 | \crt{u\uparrow} \crt{w\downarrow} \dst{t\downarrow} \dst{v\uparrow} | \Phi_0 \rangle = S^{\uparrow\downarrow}_{(wu),(tv)} \quad, \\
\end{split}
\end{equation}

where

\begin{equation}
\begin{split}
w^\prime &= \mbox{min}(u,w) \quad,\quad 
u^\prime  = \mbox{max}(u,w) \quad, \\
t^\prime &= \mbox{min}(t,v) \quad,\quad 
v^\prime = \mbox{max}(t,v) \quad. 
\end{split}
\end{equation}
Having verified that one- and two-body density matrices are trivially related to QSE overlap matrices, we will show that 
transition matrix elements can be expressed in terms of one- and two-body density matrices. 
Indeed,
\begin{widetext}
\begin{equation}
\label{eq:omega_1}
\begin{split}
\Omega^\sigma_{\lambda,a} 
&= \langle \Phi^{(\sigma)}_\lambda | \hat{O}^{(1)}_{a,\sigma} | \Phi_0 \rangle = \sum_u \omega_{u\lambda}^\sigma \langle \Phi_0 | \crt{u\sigma} \hat{O}^{(1)}_{a,\sigma} | \Phi_0 \rangle 
= \sum_u \omega_{u\lambda}^\sigma \Bigg[ \sum_{t} h_{at}^\prime \, \rho^\sigma_{ut}  + \sum_{ \substack{twv \\ \tau} } v_{atwv} \, \rho^{\sigma\tau}_{utwv} \Bigg] \\
\Omega^{\sigma\sigma}_{\lambda,ab} 
&= \langle \Phi^{(\sigma\sigma)}_\lambda | \hat{O}^{(2)}_{ab,\sigma} | \Phi_0 \rangle 
= \sum_{u<v} \omega_{uv,\lambda}^\sigma \langle \Phi_0 | \crt{v\sigma} \crt{u\sigma} \hat{O}^{(2)}_{ab,\sigma} | \Phi_0 \rangle 
= \sum_{u<v} \omega_{uv,\lambda}^\sigma \, 
\left[ \sum_{wt} v_{atbw} \, \rho^{\sigma\sigma}_{vtuw} \right] \quad, \\
\Omega^{\uparrow\downarrow}_{\lambda,ab} 
&= \langle \Phi^{(\uparrow\downarrow)}_\lambda | \hat{O}^{(3)}_{ab} | \Phi_0 \rangle 
= \sum_{uv} \omega_{uv,\lambda}^\sigma \langle \Phi_0 | \crt{v\downarrow} \crt{u\uparrow} \hat{O}^{(3)}_{ab} | \Phi_0 \rangle 
= \sum_{uv} \omega_{uv,\lambda}^\sigma \, \left[ \sum_{wt} v_{atbw} \, \rho^{\uparrow\downarrow}_{vtuw} \right] \quad . 
\end{split}
\end{equation}
\end{widetext}
The cost of computing the terms of Eq.~\eqref{eq:omega_1} is respectively of $\mathcal{O}(N_{ext} N_{act}^3 + N_{ext} N_{act} N_{qse})$,
and  $\mathcal{O}(N^2_{ext} N_{act}^4 + N^2_{ext} N^2_{act} N_{qse})$ operations.

\subsection{$R_y$ variational form}
\label{app:ry}

The quantum circuit defining the variational form used in the present work is shown in Fig.~\ref{fig:ry}.

The initial state $|x_0 x_1 x_2 x_3 x_4 x_5 \rangle$ is the computational basis state (i.e. a tensor product of eigenstates of the $Z$ Pauli operator)
representing the Hartree-Fock state in presence of tapering. For the anion and radical, this is respectively $(x_0,x_1,x_2,x_3,x_4,x_5)=(1,0,1,1,1,0)$
and $(x_0,x_1,x_2,x_3,x_4,x_5)=(1,0,1,1,0,0)$.
Observables such as the active-space Hamiltonian $\hat{H}_{act}$ and the QSE overlap and Hamiltonian operators Eq.~\eqref{eq:nevpt7} are represented as linear combinations of Pauli operators $P = \otimes_{i=0}^{n_q-1} \sigma_i$ with standard mappings \cite{bravyi2017tapering,aleksandrowicz2019qiskit}.

At the end of the circuit, Pauli operators $P$ are measured, and the results of these measurements are used to compute expectation values of relevant operators using standard techniques \cite{aleksandrowicz2019qiskit}.


\begin{thebibliography}{72}%
\makeatletter
\providecommand \@ifxundefined [1]{%
 \@ifx{#1\undefined}
}%
\providecommand \@ifnum [1]{%
 \ifnum #1\expandafter \@firstoftwo
 \else \expandafter \@secondoftwo
 \fi
}%
\providecommand \@ifx [1]{%
 \ifx #1\expandafter \@firstoftwo
 \else \expandafter \@secondoftwo
 \fi
}%
\providecommand \natexlab [1]{#1}%
\providecommand \enquote  [1]{``#1''}%
\providecommand \bibnamefont  [1]{#1}%
\providecommand \bibfnamefont [1]{#1}%
\providecommand \citenamefont [1]{#1}%
\providecommand \href@noop [0]{\@secondoftwo}%
\providecommand \href [0]{\begingroup \@sanitize@url \@href}%
\providecommand \@href[1]{\@@startlink{#1}\@@href}%
\providecommand \@@href[1]{\endgroup#1\@@endlink}%
\providecommand \@sanitize@url [0]{\catcode `\\12\catcode `\$12\catcode
  `\&12\catcode `\#12\catcode `\^12\catcode `\_12\catcode `\%12\relax}%
\providecommand \@@startlink[1]{}%
\providecommand \@@endlink[0]{}%
\providecommand \url  [0]{\begingroup\@sanitize@url \@url }%
\providecommand \@url [1]{\endgroup\@href {#1}{\urlprefix }}%
\providecommand \urlprefix  [0]{URL }%
\providecommand \Eprint [0]{\href }%
\providecommand \doibase [0]{https://doi.org/}%
\providecommand \selectlanguage [0]{\@gobble}%
\providecommand \bibinfo  [0]{\@secondoftwo}%
\providecommand \bibfield  [0]{\@secondoftwo}%
\providecommand \translation [1]{[#1]}%
\providecommand \BibitemOpen [0]{}%
\providecommand \bibitemStop [0]{}%
\providecommand \bibitemNoStop [0]{.\EOS\space}%
\providecommand \EOS [0]{\spacefactor3000\relax}%
\providecommand \BibitemShut  [1]{\csname bibitem#1\endcsname}%
\let\auto@bib@innerbib\@empty
\bibitem [{\citenamefont {Georgescu}\ \emph {et~al.}(2014)\citenamefont
  {Georgescu}, \citenamefont {Ashhab},\ and\ \citenamefont
  {Nori}}]{georgescu2014quantum}%
  \BibitemOpen
  \bibfield  {author} {\bibinfo {author} {\bibfnamefont {I.~M.}\ \bibnamefont
  {Georgescu}}, \bibinfo {author} {\bibfnamefont {S.}~\bibnamefont {Ashhab}},\
  and\ \bibinfo {author} {\bibfnamefont {F.}~\bibnamefont {Nori}},\ }\href
  {https://journals.aps.org/rmp/abstract/10.1103/RevModPhys.86.153} {\bibfield
  {journal} {\bibinfo  {journal} {Rev. Mod. Phys}\ }\textbf {\bibinfo {volume}
  {86}},\ \bibinfo {pages} {153} (\bibinfo {year} {2014})}\BibitemShut
  {NoStop}%
\bibitem [{\citenamefont {Cao}\ \emph {et~al.}(2019)\citenamefont {Cao},
  \citenamefont {Romero}, \citenamefont {Olson}, \citenamefont {Degroote},
  \citenamefont {Johnson}, \citenamefont {Kieferov{\'a}}, \citenamefont
  {Kivlichan}, \citenamefont {Menke}, \citenamefont {Peropadre}, \citenamefont
  {Sawaya} \emph {et~al.}}]{cao2019quantum}%
  \BibitemOpen
  \bibfield  {author} {\bibinfo {author} {\bibfnamefont {Y.}~\bibnamefont
  {Cao}}, \bibinfo {author} {\bibfnamefont {J.}~\bibnamefont {Romero}},
  \bibinfo {author} {\bibfnamefont {J.~P.}\ \bibnamefont {Olson}}, \bibinfo
  {author} {\bibfnamefont {M.}~\bibnamefont {Degroote}}, \bibinfo {author}
  {\bibfnamefont {P.~D.}\ \bibnamefont {Johnson}}, \bibinfo {author}
  {\bibfnamefont {M.}~\bibnamefont {Kieferov{\'a}}}, \bibinfo {author}
  {\bibfnamefont {I.~D.}\ \bibnamefont {Kivlichan}}, \bibinfo {author}
  {\bibfnamefont {T.}~\bibnamefont {Menke}}, \bibinfo {author} {\bibfnamefont
  {B.}~\bibnamefont {Peropadre}}, \bibinfo {author} {\bibfnamefont {N.~P.}\
  \bibnamefont {Sawaya}}, \emph {et~al.},\ }\href
  {https://pubs.acs.org/doi/10.1021/acs.chemrev.8b00803} {\bibfield  {journal}
  {\bibinfo  {journal} {Chem. Rev}\ }\textbf {\bibinfo {volume} {119}},\
  \bibinfo {pages} {10856} (\bibinfo {year} {2019})}\BibitemShut {NoStop}%
\bibitem [{\citenamefont {Cerezo}\ \emph {et~al.}(2021)\citenamefont {Cerezo},
  \citenamefont {Arrasmith}, \citenamefont {Babbush}, \citenamefont {Benjamin},
  \citenamefont {Endo}, \citenamefont {Fujii}, \citenamefont {McClean},
  \citenamefont {Mitarai}, \citenamefont {Yuan}, \citenamefont {Cincio} \emph
  {et~al.}}]{cerezo2020variational}%
  \BibitemOpen
  \bibfield  {author} {\bibinfo {author} {\bibfnamefont {M.}~\bibnamefont
  {Cerezo}}, \bibinfo {author} {\bibfnamefont {A.}~\bibnamefont {Arrasmith}},
  \bibinfo {author} {\bibfnamefont {R.}~\bibnamefont {Babbush}}, \bibinfo
  {author} {\bibfnamefont {S.~C.}\ \bibnamefont {Benjamin}}, \bibinfo {author}
  {\bibfnamefont {S.}~\bibnamefont {Endo}}, \bibinfo {author} {\bibfnamefont
  {K.}~\bibnamefont {Fujii}}, \bibinfo {author} {\bibfnamefont {J.~R.}\
  \bibnamefont {McClean}}, \bibinfo {author} {\bibfnamefont {K.}~\bibnamefont
  {Mitarai}}, \bibinfo {author} {\bibfnamefont {X.}~\bibnamefont {Yuan}},
  \bibinfo {author} {\bibfnamefont {L.}~\bibnamefont {Cincio}}, \emph
  {et~al.},\ }\href {https://www.nature.com/articles/s42254-021-00348-9}
  {\bibfield  {journal} {\bibinfo  {journal} {Nat. Rev. Phys}\ ,\ \bibinfo
  {pages} {1}} (\bibinfo {year} {2021})}\BibitemShut {NoStop}%
\bibitem [{\citenamefont {Bauer}\ \emph {et~al.}(2020)\citenamefont {Bauer},
  \citenamefont {Bravyi}, \citenamefont {Motta},\ and\ \citenamefont
  {Kin-Lic~Chan}}]{bauer2020quantum}%
  \BibitemOpen
  \bibfield  {author} {\bibinfo {author} {\bibfnamefont {B.}~\bibnamefont
  {Bauer}}, \bibinfo {author} {\bibfnamefont {S.}~\bibnamefont {Bravyi}},
  \bibinfo {author} {\bibfnamefont {M.}~\bibnamefont {Motta}},\ and\ \bibinfo
  {author} {\bibfnamefont {G.}~\bibnamefont {Kin-Lic~Chan}},\ }\href
  {https://pubs.acs.org/doi/10.1021/acs.chemrev.9b00829} {\bibfield  {journal}
  {\bibinfo  {journal} {Chem. Rev}\ }\textbf {\bibinfo {volume} {120}},\
  \bibinfo {pages} {12685} (\bibinfo {year} {2020})}\BibitemShut {NoStop}%
\bibitem [{\citenamefont {McArdle}\ \emph {et~al.}(2020)\citenamefont
  {McArdle}, \citenamefont {Endo}, \citenamefont {Aspuru-Guzik}, \citenamefont
  {Benjamin},\ and\ \citenamefont {Yuan}}]{mcardle2020quantum}%
  \BibitemOpen
  \bibfield  {author} {\bibinfo {author} {\bibfnamefont {S.}~\bibnamefont
  {McArdle}}, \bibinfo {author} {\bibfnamefont {S.}~\bibnamefont {Endo}},
  \bibinfo {author} {\bibfnamefont {A.}~\bibnamefont {Aspuru-Guzik}}, \bibinfo
  {author} {\bibfnamefont {S.~C.}\ \bibnamefont {Benjamin}},\ and\ \bibinfo
  {author} {\bibfnamefont {X.}~\bibnamefont {Yuan}},\ }\href
  {https://doi.org/10.1103/RevModPhys.92.015003} {\bibfield  {journal}
  {\bibinfo  {journal} {Rev. Mod. Phys}\ }\textbf {\bibinfo {volume} {92}},\
  \bibinfo {pages} {015003} (\bibinfo {year} {2020})}\BibitemShut {NoStop}%
\bibitem [{\citenamefont {Motta}\ and\ \citenamefont
  {Rice}(2021)}]{motta2021emerging}%
  \BibitemOpen
  \bibfield  {author} {\bibinfo {author} {\bibfnamefont {M.}~\bibnamefont
  {Motta}}\ and\ \bibinfo {author} {\bibfnamefont {J.~E.}\ \bibnamefont
  {Rice}},\ }\href
  {https://wires.onlinelibrary.wiley.com/doi/abs/10.1002/wcms.1580} {\bibfield
  {journal} {\bibinfo  {journal} {WIREs Comput. Mol. Sci}\ ,\ \bibinfo {pages}
  {e1580}} (\bibinfo {year} {2021})}\BibitemShut {NoStop}%
\bibitem [{\citenamefont {Friesner}(2005)}]{friesner2005ab}%
  \BibitemOpen
  \bibfield  {author} {\bibinfo {author} {\bibfnamefont {R.~A.}\ \bibnamefont
  {Friesner}},\ }\href {https://www.pnas.org/content/102/19/6648.short}
  {\bibfield  {journal} {\bibinfo  {journal} {Proc. Nat. Acad. Sci. USA}\
  }\textbf {\bibinfo {volume} {102}},\ \bibinfo {pages} {6648} (\bibinfo {year}
  {2005})}\BibitemShut {NoStop}%
\bibitem [{\citenamefont {Helgaker}\ \emph {et~al.}(2012)\citenamefont
  {Helgaker}, \citenamefont {Coriani}, \citenamefont {J{\o}rgensen},
  \citenamefont {Kristensen}, \citenamefont {Olsen},\ and\ \citenamefont
  {Ruud}}]{helgaker2012recent}%
  \BibitemOpen
  \bibfield  {author} {\bibinfo {author} {\bibfnamefont {T.}~\bibnamefont
  {Helgaker}}, \bibinfo {author} {\bibfnamefont {S.}~\bibnamefont {Coriani}},
  \bibinfo {author} {\bibfnamefont {P.}~\bibnamefont {J{\o}rgensen}}, \bibinfo
  {author} {\bibfnamefont {K.}~\bibnamefont {Kristensen}}, \bibinfo {author}
  {\bibfnamefont {J.}~\bibnamefont {Olsen}},\ and\ \bibinfo {author}
  {\bibfnamefont {K.}~\bibnamefont {Ruud}},\ }\href
  {https://doi.org/10.1021/cr2002239} {\bibfield  {journal} {\bibinfo
  {journal} {Chem. Rev}\ }\textbf {\bibinfo {volume} {112}},\ \bibinfo {pages}
  {543} (\bibinfo {year} {2012})}\BibitemShut {NoStop}%
\bibitem [{\citenamefont {Helgaker}\ \emph {et~al.}(2014)\citenamefont
  {Helgaker}, \citenamefont {Jorgensen},\ and\ \citenamefont
  {Olsen}}]{helgaker2014molecular}%
  \BibitemOpen
  \bibfield  {author} {\bibinfo {author} {\bibfnamefont {T.}~\bibnamefont
  {Helgaker}}, \bibinfo {author} {\bibfnamefont {P.}~\bibnamefont
  {Jorgensen}},\ and\ \bibinfo {author} {\bibfnamefont {J.}~\bibnamefont
  {Olsen}},\ }\href@noop {} {\emph {\bibinfo {title} {Molecular
  electronic-structure theory}}}\ (\bibinfo  {publisher} {John Wiley \& Sons},\
  \bibinfo {year} {2014})\BibitemShut {NoStop}%
\bibitem [{\citenamefont {Bravyi}\ \emph {et~al.}(2016)\citenamefont {Bravyi},
  \citenamefont {Smith},\ and\ \citenamefont {Smolin}}]{bravyi2016trading}%
  \BibitemOpen
  \bibfield  {author} {\bibinfo {author} {\bibfnamefont {S.}~\bibnamefont
  {Bravyi}}, \bibinfo {author} {\bibfnamefont {G.}~\bibnamefont {Smith}},\ and\
  \bibinfo {author} {\bibfnamefont {J.~A.}\ \bibnamefont {Smolin}},\ }\href
  {https://journals.aps.org/prx/abstract/10.1103/PhysRevX.6.021043} {\bibfield
  {journal} {\bibinfo  {journal} {Phys. Rev. X}\ }\textbf {\bibinfo {volume}
  {6}},\ \bibinfo {pages} {021043} (\bibinfo {year} {2016})}\BibitemShut
  {NoStop}%
\bibitem [{\citenamefont {Kreula}\ \emph {et~al.}(2016)\citenamefont {Kreula},
  \citenamefont {Garc{\'\i}a-{\'A}lvarez}, \citenamefont {Lamata},
  \citenamefont {Clark}, \citenamefont {Solano},\ and\ \citenamefont
  {Jaksch}}]{kreula2016few}%
  \BibitemOpen
  \bibfield  {author} {\bibinfo {author} {\bibfnamefont {J.~M.}\ \bibnamefont
  {Kreula}}, \bibinfo {author} {\bibfnamefont {L.}~\bibnamefont
  {Garc{\'\i}a-{\'A}lvarez}}, \bibinfo {author} {\bibfnamefont
  {L.}~\bibnamefont {Lamata}}, \bibinfo {author} {\bibfnamefont {S.~R.}\
  \bibnamefont {Clark}}, \bibinfo {author} {\bibfnamefont {E.}~\bibnamefont
  {Solano}},\ and\ \bibinfo {author} {\bibfnamefont {D.}~\bibnamefont
  {Jaksch}},\ }\href
  {https://epjquantumtechnology.springeropen.com/articles/10.1140/epjqt/s40507-016-0049-1}
  {\bibfield  {journal} {\bibinfo  {journal} {EPJ Quantum Technology}\ }\textbf
  {\bibinfo {volume} {3}},\ \bibinfo {pages} {1} (\bibinfo {year}
  {2016})}\BibitemShut {NoStop}%
\bibitem [{\citenamefont {Yamazaki}\ \emph {et~al.}(2018)\citenamefont
  {Yamazaki}, \citenamefont {Matsuura}, \citenamefont {Narimani}, \citenamefont
  {Saidmuradov},\ and\ \citenamefont {Zaribafiyan}}]{yamazaki2018towards}%
  \BibitemOpen
  \bibfield  {author} {\bibinfo {author} {\bibfnamefont {T.}~\bibnamefont
  {Yamazaki}}, \bibinfo {author} {\bibfnamefont {S.}~\bibnamefont {Matsuura}},
  \bibinfo {author} {\bibfnamefont {A.}~\bibnamefont {Narimani}}, \bibinfo
  {author} {\bibfnamefont {A.}~\bibnamefont {Saidmuradov}},\ and\ \bibinfo
  {author} {\bibfnamefont {A.}~\bibnamefont {Zaribafiyan}},\ }\href
  {https://arxiv.org/abs/1806.01305} {\bibfield  {journal} {\bibinfo  {journal}
  {arXiv:1806.01305}\ } (\bibinfo {year} {2018})}\BibitemShut {NoStop}%
\bibitem [{\citenamefont {Peng}\ \emph {et~al.}(2020)\citenamefont {Peng},
  \citenamefont {Harrow}, \citenamefont {Ozols},\ and\ \citenamefont
  {Wu}}]{peng2020simulating}%
  \BibitemOpen
  \bibfield  {author} {\bibinfo {author} {\bibfnamefont {T.}~\bibnamefont
  {Peng}}, \bibinfo {author} {\bibfnamefont {A.~W.}\ \bibnamefont {Harrow}},
  \bibinfo {author} {\bibfnamefont {M.}~\bibnamefont {Ozols}},\ and\ \bibinfo
  {author} {\bibfnamefont {X.}~\bibnamefont {Wu}},\ }\href
  {https://journals.aps.org/prl/abstract/10.1103/PhysRevLett.125.150504}
  {\bibfield  {journal} {\bibinfo  {journal} {Phys. Rev. Lett}\ }\textbf
  {\bibinfo {volume} {125}},\ \bibinfo {pages} {150504} (\bibinfo {year}
  {2020})}\BibitemShut {NoStop}%
\bibitem [{\citenamefont {Takeshita}\ \emph {et~al.}(2020)\citenamefont
  {Takeshita}, \citenamefont {Rubin}, \citenamefont {Jiang}, \citenamefont
  {Lee}, \citenamefont {Babbush},\ and\ \citenamefont
  {McClean}}]{takeshita2020increasing}%
  \BibitemOpen
  \bibfield  {author} {\bibinfo {author} {\bibfnamefont {T.}~\bibnamefont
  {Takeshita}}, \bibinfo {author} {\bibfnamefont {N.~C.}\ \bibnamefont
  {Rubin}}, \bibinfo {author} {\bibfnamefont {Z.}~\bibnamefont {Jiang}},
  \bibinfo {author} {\bibfnamefont {E.}~\bibnamefont {Lee}}, \bibinfo {author}
  {\bibfnamefont {R.}~\bibnamefont {Babbush}},\ and\ \bibinfo {author}
  {\bibfnamefont {J.~R.}\ \bibnamefont {McClean}},\ }\href
  {https://journals.aps.org/prx/pdf/10.1103/PhysRevX.10.011004} {\bibfield
  {journal} {\bibinfo  {journal} {Phys. Rev. X}\ }\textbf {\bibinfo {volume}
  {10}},\ \bibinfo {pages} {011004} (\bibinfo {year} {2020})}\BibitemShut
  {NoStop}%
\bibitem [{\citenamefont {Kawashima}\ \emph {et~al.}(2021)\citenamefont
  {Kawashima}, \citenamefont {Lloyd}, \citenamefont {Coons}, \citenamefont
  {Nam}, \citenamefont {Matsuura}, \citenamefont {Garza}, \citenamefont
  {Johri}, \citenamefont {Huntington}, \citenamefont {Senicourt}, \citenamefont
  {Maksymov} \emph {et~al.}}]{kawashima2021optimizing}%
  \BibitemOpen
  \bibfield  {author} {\bibinfo {author} {\bibfnamefont {Y.}~\bibnamefont
  {Kawashima}}, \bibinfo {author} {\bibfnamefont {E.}~\bibnamefont {Lloyd}},
  \bibinfo {author} {\bibfnamefont {M.~P.}\ \bibnamefont {Coons}}, \bibinfo
  {author} {\bibfnamefont {Y.}~\bibnamefont {Nam}}, \bibinfo {author}
  {\bibfnamefont {S.}~\bibnamefont {Matsuura}}, \bibinfo {author}
  {\bibfnamefont {A.~J.}\ \bibnamefont {Garza}}, \bibinfo {author}
  {\bibfnamefont {S.}~\bibnamefont {Johri}}, \bibinfo {author} {\bibfnamefont
  {L.}~\bibnamefont {Huntington}}, \bibinfo {author} {\bibfnamefont
  {V.}~\bibnamefont {Senicourt}}, \bibinfo {author} {\bibfnamefont {A.~O.}\
  \bibnamefont {Maksymov}}, \emph {et~al.},\ }\href
  {https://www.nature.com/articles/s42005-021-00751-9} {\bibfield  {journal}
  {\bibinfo  {journal} {Nat. Commun}\ }\textbf {\bibinfo {volume} {4}},\
  \bibinfo {pages} {1} (\bibinfo {year} {2021})}\BibitemShut {NoStop}%
\bibitem [{\citenamefont {Mitarai}\ and\ \citenamefont
  {Fujii}(2021)}]{mitarai2021constructing}%
  \BibitemOpen
  \bibfield  {author} {\bibinfo {author} {\bibfnamefont {K.}~\bibnamefont
  {Mitarai}}\ and\ \bibinfo {author} {\bibfnamefont {K.}~\bibnamefont
  {Fujii}},\ }\href
  {https://iopscience.iop.org/article/10.1088/1367-2630/abd7bc/meta} {\bibfield
   {journal} {\bibinfo  {journal} {New J. Phys}\ }\textbf {\bibinfo {volume}
  {23}},\ \bibinfo {pages} {023021} (\bibinfo {year} {2021})}\BibitemShut
  {NoStop}%
\bibitem [{\citenamefont {Yuan}\ \emph {et~al.}(2021)\citenamefont {Yuan},
  \citenamefont {Sun}, \citenamefont {Liu}, \citenamefont {Zhao},\ and\
  \citenamefont {Zhou}}]{yuan2021quantum}%
  \BibitemOpen
  \bibfield  {author} {\bibinfo {author} {\bibfnamefont {X.}~\bibnamefont
  {Yuan}}, \bibinfo {author} {\bibfnamefont {J.}~\bibnamefont {Sun}}, \bibinfo
  {author} {\bibfnamefont {J.}~\bibnamefont {Liu}}, \bibinfo {author}
  {\bibfnamefont {Q.}~\bibnamefont {Zhao}},\ and\ \bibinfo {author}
  {\bibfnamefont {Y.}~\bibnamefont {Zhou}},\ }\href
  {https://journals.aps.org/prl/abstract/10.1103/PhysRevLett.127.040501}
  {\bibfield  {journal} {\bibinfo  {journal} {Phys. Rev. Lett}\ }\textbf
  {\bibinfo {volume} {127}},\ \bibinfo {pages} {040501} (\bibinfo {year}
  {2021})}\BibitemShut {NoStop}%
\bibitem [{\citenamefont {Eddins}\ \emph {et~al.}(2022)\citenamefont {Eddins},
  \citenamefont {Motta}, \citenamefont {Gujarati}, \citenamefont {Bravyi},
  \citenamefont {Mezzacapo}, \citenamefont {Hadfield},\ and\ \citenamefont
  {Sheldon}}]{eddins2021doubling}%
  \BibitemOpen
  \bibfield  {author} {\bibinfo {author} {\bibfnamefont {A.}~\bibnamefont
  {Eddins}}, \bibinfo {author} {\bibfnamefont {M.}~\bibnamefont {Motta}},
  \bibinfo {author} {\bibfnamefont {T.~P.}\ \bibnamefont {Gujarati}}, \bibinfo
  {author} {\bibfnamefont {S.}~\bibnamefont {Bravyi}}, \bibinfo {author}
  {\bibfnamefont {A.}~\bibnamefont {Mezzacapo}}, \bibinfo {author}
  {\bibfnamefont {C.}~\bibnamefont {Hadfield}},\ and\ \bibinfo {author}
  {\bibfnamefont {S.}~\bibnamefont {Sheldon}},\ }\href
  {https://doi.org/10.1103/PRXQuantum.3.010309} {\bibfield  {journal} {\bibinfo
   {journal} {Phys. Rev. X Quantum}\ }\textbf {\bibinfo {volume} {3}},\
  \bibinfo {pages} {010309} (\bibinfo {year} {2022})}\BibitemShut {NoStop}%
\bibitem [{\citenamefont {Farhi}\ \emph {et~al.}(2014)\citenamefont {Farhi},
  \citenamefont {Goldstone},\ and\ \citenamefont {Gutmann}}]{farhi2014quantum}%
  \BibitemOpen
  \bibfield  {author} {\bibinfo {author} {\bibfnamefont {E.}~\bibnamefont
  {Farhi}}, \bibinfo {author} {\bibfnamefont {J.}~\bibnamefont {Goldstone}},\
  and\ \bibinfo {author} {\bibfnamefont {S.}~\bibnamefont {Gutmann}},\ }\href
  {https://arxiv.org/abs/1411.4028} {\bibfield  {journal} {\bibinfo  {journal}
  {arXiv:1411.4028}\ } (\bibinfo {year} {2014})}\BibitemShut {NoStop}%
\bibitem [{\citenamefont {Peruzzo}\ \emph {et~al.}(2014)\citenamefont
  {Peruzzo}, \citenamefont {McClean}, \citenamefont {Shadbolt}, \citenamefont
  {Yung}, \citenamefont {Zhou}, \citenamefont {Love}, \citenamefont
  {Aspuru-Guzik},\ and\ \citenamefont {O'Brien}}]{peruzzo2014variational}%
  \BibitemOpen
  \bibfield  {author} {\bibinfo {author} {\bibfnamefont {A.}~\bibnamefont
  {Peruzzo}}, \bibinfo {author} {\bibfnamefont {J.}~\bibnamefont {McClean}},
  \bibinfo {author} {\bibfnamefont {P.}~\bibnamefont {Shadbolt}}, \bibinfo
  {author} {\bibfnamefont {M.-H.}\ \bibnamefont {Yung}}, \bibinfo {author}
  {\bibfnamefont {X.-Q.}\ \bibnamefont {Zhou}}, \bibinfo {author}
  {\bibfnamefont {P.~J.}\ \bibnamefont {Love}}, \bibinfo {author}
  {\bibfnamefont {A.}~\bibnamefont {Aspuru-Guzik}},\ and\ \bibinfo {author}
  {\bibfnamefont {J.~L.}\ \bibnamefont {O'Brien}},\ }\href
  {https://doi.org/10.1038/ncomms5213} {\bibfield  {journal} {\bibinfo
  {journal} {Nat. Commun}\ }\textbf {\bibinfo {volume} {5}},\ \bibinfo {pages}
  {4213} (\bibinfo {year} {2014})}\BibitemShut {NoStop}%
\bibitem [{\citenamefont {McClean}\ \emph {et~al.}(2016)\citenamefont
  {McClean}, \citenamefont {Romero}, \citenamefont {Babbush},\ and\
  \citenamefont {Aspuru-Guzik}}]{mcclean2016theory}%
  \BibitemOpen
  \bibfield  {author} {\bibinfo {author} {\bibfnamefont {J.~R.}\ \bibnamefont
  {McClean}}, \bibinfo {author} {\bibfnamefont {J.}~\bibnamefont {Romero}},
  \bibinfo {author} {\bibfnamefont {R.}~\bibnamefont {Babbush}},\ and\ \bibinfo
  {author} {\bibfnamefont {A.}~\bibnamefont {Aspuru-Guzik}},\ }\href
  {https://iopscience.iop.org/article/10.1088/1367-2630/18/2/023023} {\bibfield
   {journal} {\bibinfo  {journal} {New J. Phys}\ }\textbf {\bibinfo {volume}
  {18}},\ \bibinfo {pages} {023023} (\bibinfo {year} {2016})}\BibitemShut
  {NoStop}%
\bibitem [{\citenamefont {Romero}\ \emph {et~al.}(2018)\citenamefont {Romero},
  \citenamefont {Babbush}, \citenamefont {McClean}, \citenamefont {Hempel},
  \citenamefont {Love},\ and\ \citenamefont
  {Aspuru-Guzik}}]{romero2018strategies}%
  \BibitemOpen
  \bibfield  {author} {\bibinfo {author} {\bibfnamefont {J.}~\bibnamefont
  {Romero}}, \bibinfo {author} {\bibfnamefont {R.}~\bibnamefont {Babbush}},
  \bibinfo {author} {\bibfnamefont {J.~R.}\ \bibnamefont {McClean}}, \bibinfo
  {author} {\bibfnamefont {C.}~\bibnamefont {Hempel}}, \bibinfo {author}
  {\bibfnamefont {P.~J.}\ \bibnamefont {Love}},\ and\ \bibinfo {author}
  {\bibfnamefont {A.}~\bibnamefont {Aspuru-Guzik}},\ }\href
  {https://iopscience.iop.org/article/10.1088/2058-9565/aad3e4/meta} {\bibfield
   {journal} {\bibinfo  {journal} {Quantum Sci. Technol.}\ }\textbf {\bibinfo
  {volume} {4}},\ \bibinfo {pages} {014008} (\bibinfo {year}
  {2018})}\BibitemShut {NoStop}%
\bibitem [{\citenamefont {McClean}\ \emph {et~al.}(2017)\citenamefont
  {McClean}, \citenamefont {Kimchi-Schwartz}, \citenamefont {Carter},\ and\
  \citenamefont {De~Jong}}]{mcclean2017hybrid}%
  \BibitemOpen
  \bibfield  {author} {\bibinfo {author} {\bibfnamefont {J.~R.}\ \bibnamefont
  {McClean}}, \bibinfo {author} {\bibfnamefont {M.~E.}\ \bibnamefont
  {Kimchi-Schwartz}}, \bibinfo {author} {\bibfnamefont {J.}~\bibnamefont
  {Carter}},\ and\ \bibinfo {author} {\bibfnamefont {W.~A.}\ \bibnamefont
  {De~Jong}},\ }\href
  {https://journals.aps.org/pra/abstract/10.1103/PhysRevA.95.042308} {\bibfield
   {journal} {\bibinfo  {journal} {Phys. Rev. A}\ }\textbf {\bibinfo {volume}
  {95}},\ \bibinfo {pages} {042308} (\bibinfo {year} {2017})}\BibitemShut
  {NoStop}%
\bibitem [{\citenamefont {Colless}\ \emph {et~al.}(2018)\citenamefont
  {Colless}, \citenamefont {Ramasesh}, \citenamefont {Dahlen}, \citenamefont
  {Blok}, \citenamefont {Kimchi-Schwartz}, \citenamefont {McClean},
  \citenamefont {Carter}, \citenamefont {De~Jong},\ and\ \citenamefont
  {Siddiqi}}]{colless2018computation}%
  \BibitemOpen
  \bibfield  {author} {\bibinfo {author} {\bibfnamefont {J.~I.}\ \bibnamefont
  {Colless}}, \bibinfo {author} {\bibfnamefont {V.~V.}\ \bibnamefont
  {Ramasesh}}, \bibinfo {author} {\bibfnamefont {D.}~\bibnamefont {Dahlen}},
  \bibinfo {author} {\bibfnamefont {M.~S.}\ \bibnamefont {Blok}}, \bibinfo
  {author} {\bibfnamefont {M.}~\bibnamefont {Kimchi-Schwartz}}, \bibinfo
  {author} {\bibfnamefont {J.}~\bibnamefont {McClean}}, \bibinfo {author}
  {\bibfnamefont {J.}~\bibnamefont {Carter}}, \bibinfo {author} {\bibfnamefont
  {W.}~\bibnamefont {De~Jong}},\ and\ \bibinfo {author} {\bibfnamefont
  {I.}~\bibnamefont {Siddiqi}},\ }\href
  {https://journals.aps.org/prx/abstract/10.1103/PhysRevX.8.011021} {\bibfield
  {journal} {\bibinfo  {journal} {Phys. Rev. X}\ }\textbf {\bibinfo {volume}
  {8}},\ \bibinfo {pages} {011021} (\bibinfo {year} {2018})}\BibitemShut
  {NoStop}%
\bibitem [{\citenamefont {Huggins}\ \emph {et~al.}(2020)\citenamefont
  {Huggins}, \citenamefont {Lee}, \citenamefont {Baek}, \citenamefont
  {O’Gorman},\ and\ \citenamefont {Whaley}}]{huggins2020non}%
  \BibitemOpen
  \bibfield  {author} {\bibinfo {author} {\bibfnamefont {W.~J.}\ \bibnamefont
  {Huggins}}, \bibinfo {author} {\bibfnamefont {J.}~\bibnamefont {Lee}},
  \bibinfo {author} {\bibfnamefont {U.}~\bibnamefont {Baek}}, \bibinfo {author}
  {\bibfnamefont {B.}~\bibnamefont {O’Gorman}},\ and\ \bibinfo {author}
  {\bibfnamefont {K.~B.}\ \bibnamefont {Whaley}},\ }\href
  {https://iopscience.iop.org/article/10.1088/1367-2630/ab867b} {\bibfield
  {journal} {\bibinfo  {journal} {New J. Phys}\ }\textbf {\bibinfo {volume}
  {22}},\ \bibinfo {pages} {073009} (\bibinfo {year} {2020})}\BibitemShut
  {NoStop}%
\bibitem [{\citenamefont {Angeli}\ \emph
  {et~al.}(2001{\natexlab{a}})\citenamefont {Angeli}, \citenamefont
  {Cimiraglia}, \citenamefont {Evangelisti}, \citenamefont {Leininger},\ and\
  \citenamefont {Malrieu}}]{angeli2001introduction}%
  \BibitemOpen
  \bibfield  {author} {\bibinfo {author} {\bibfnamefont {C.}~\bibnamefont
  {Angeli}}, \bibinfo {author} {\bibfnamefont {R.}~\bibnamefont {Cimiraglia}},
  \bibinfo {author} {\bibfnamefont {S.}~\bibnamefont {Evangelisti}}, \bibinfo
  {author} {\bibfnamefont {T.}~\bibnamefont {Leininger}},\ and\ \bibinfo
  {author} {\bibfnamefont {J.-P.}\ \bibnamefont {Malrieu}},\ }\href
  {https://aip.scitation.org/doi/10.1063/1.1361246} {\bibfield  {journal}
  {\bibinfo  {journal} {J. Chem. Phys}\ }\textbf {\bibinfo {volume} {114}},\
  \bibinfo {pages} {10252} (\bibinfo {year} {2001}{\natexlab{a}})}\BibitemShut
  {NoStop}%
\bibitem [{\citenamefont {Angeli}\ \emph
  {et~al.}(2001{\natexlab{b}})\citenamefont {Angeli}, \citenamefont
  {Cimiraglia},\ and\ \citenamefont {Malrieu}}]{angeli2001n}%
  \BibitemOpen
  \bibfield  {author} {\bibinfo {author} {\bibfnamefont {C.}~\bibnamefont
  {Angeli}}, \bibinfo {author} {\bibfnamefont {R.}~\bibnamefont {Cimiraglia}},\
  and\ \bibinfo {author} {\bibfnamefont {J.-P.}\ \bibnamefont {Malrieu}},\
  }\href
  {https://www.sciencedirect.com/science/article/abs/pii/S0009261401013033}
  {\bibfield  {journal} {\bibinfo  {journal} {Chem. Phys. Lett}\ }\textbf
  {\bibinfo {volume} {350}},\ \bibinfo {pages} {297} (\bibinfo {year}
  {2001}{\natexlab{b}})}\BibitemShut {NoStop}%
\bibitem [{\citenamefont {Sokolov}\ and\ \citenamefont
  {Chan}(2016)}]{sokolov2016time}%
  \BibitemOpen
  \bibfield  {author} {\bibinfo {author} {\bibfnamefont {A.~Y.}\ \bibnamefont
  {Sokolov}}\ and\ \bibinfo {author} {\bibfnamefont {G.~K.-L.}\ \bibnamefont
  {Chan}},\ }\href {https://aip.scitation.org/doi/10.1063/1.4941606} {\bibfield
   {journal} {\bibinfo  {journal} {J. Chem. Phys}\ }\textbf {\bibinfo {volume}
  {144}},\ \bibinfo {pages} {064102} (\bibinfo {year} {2016})}\BibitemShut
  {NoStop}%
\bibitem [{\citenamefont {Sokolov}\ \emph {et~al.}(2017)\citenamefont
  {Sokolov}, \citenamefont {Guo}, \citenamefont {Ronca},\ and\ \citenamefont
  {Chan}}]{sokolov2017time}%
  \BibitemOpen
  \bibfield  {author} {\bibinfo {author} {\bibfnamefont {A.~Y.}\ \bibnamefont
  {Sokolov}}, \bibinfo {author} {\bibfnamefont {S.}~\bibnamefont {Guo}},
  \bibinfo {author} {\bibfnamefont {E.}~\bibnamefont {Ronca}},\ and\ \bibinfo
  {author} {\bibfnamefont {G.~K.-L.}\ \bibnamefont {Chan}},\ }\href
  {https://aip.scitation.org/doi/10.1063/1.4986975} {\bibfield  {journal}
  {\bibinfo  {journal} {J. Chem. Phys}\ }\textbf {\bibinfo {volume} {146}},\
  \bibinfo {pages} {244102} (\bibinfo {year} {2017})}\BibitemShut {NoStop}%
\bibitem [{\citenamefont {Knizia}(2013)}]{knizia2013intrinsic}%
  \BibitemOpen
  \bibfield  {author} {\bibinfo {author} {\bibfnamefont {G.}~\bibnamefont
  {Knizia}},\ }\href {https://pubs.acs.org/doi/abs/10.1021/ct400687b}
  {\bibfield  {journal} {\bibinfo  {journal} {J. Chem. Theory Comput}\ }\textbf
  {\bibinfo {volume} {9}},\ \bibinfo {pages} {4834} (\bibinfo {year}
  {2013})}\BibitemShut {NoStop}%
\bibitem [{\citenamefont {Senjean}\ \emph {et~al.}(2021)\citenamefont
  {Senjean}, \citenamefont {Sen}, \citenamefont {Repisky}, \citenamefont
  {Knizia},\ and\ \citenamefont {Visscher}}]{senjean2020generalization}%
  \BibitemOpen
  \bibfield  {author} {\bibinfo {author} {\bibfnamefont {B.}~\bibnamefont
  {Senjean}}, \bibinfo {author} {\bibfnamefont {S.}~\bibnamefont {Sen}},
  \bibinfo {author} {\bibfnamefont {M.}~\bibnamefont {Repisky}}, \bibinfo
  {author} {\bibfnamefont {G.}~\bibnamefont {Knizia}},\ and\ \bibinfo {author}
  {\bibfnamefont {L.}~\bibnamefont {Visscher}},\ }\href
  {https://pubs.acs.org/doi/10.1021/acs.jctc.0c00964} {\bibfield  {journal}
  {\bibinfo  {journal} {J. Chem. Theory Comput}\ }\textbf {\bibinfo {volume}
  {17}},\ \bibinfo {pages} {1337} (\bibinfo {year} {2021})}\BibitemShut
  {NoStop}%
\bibitem [{\citenamefont {Schwilk}\ \emph {et~al.}(2017)\citenamefont
  {Schwilk}, \citenamefont {Ma}, \citenamefont {Köppl},\ and\ \citenamefont
  {Werner}}]{schwilkIAO}%
  \BibitemOpen
  \bibfield  {author} {\bibinfo {author} {\bibfnamefont {M.}~\bibnamefont
  {Schwilk}}, \bibinfo {author} {\bibfnamefont {Q.}~\bibnamefont {Ma}},
  \bibinfo {author} {\bibfnamefont {C.}~\bibnamefont {Köppl}},\ and\ \bibinfo
  {author} {\bibfnamefont {H.-J.}\ \bibnamefont {Werner}},\ }\href
  {https://doi.org/10.1021/acs.jctc.7b00554} {\bibfield  {journal} {\bibinfo
  {journal} {J. Chem. Theory Comput}\ }\textbf {\bibinfo {volume} {13}},\
  \bibinfo {pages} {3650} (\bibinfo {year} {2017})}\BibitemShut {NoStop}%
\bibitem [{\citenamefont {Manz}\ and\ \citenamefont {Limas}(2016)}]{ManzIAO}%
  \BibitemOpen
  \bibfield  {author} {\bibinfo {author} {\bibfnamefont {T.~A.}\ \bibnamefont
  {Manz}}\ and\ \bibinfo {author} {\bibfnamefont {N.~G.}\ \bibnamefont
  {Limas}},\ }\href {https://doi.org/10.1039/C6RA04656H} {\bibfield  {journal}
  {\bibinfo  {journal} {RSC Adv.}\ }\textbf {\bibinfo {volume} {6}},\ \bibinfo
  {pages} {47771} (\bibinfo {year} {2016})}\BibitemShut {NoStop}%
\bibitem [{\citenamefont {West}\ \emph {et~al.}(2013)\citenamefont {West},
  \citenamefont {Schmidt}, \citenamefont {Gordon},\ and\ \citenamefont
  {Ruedenberg}}]{WestIAO}%
  \BibitemOpen
  \bibfield  {author} {\bibinfo {author} {\bibfnamefont {A.~C.}\ \bibnamefont
  {West}}, \bibinfo {author} {\bibfnamefont {M.~W.}\ \bibnamefont {Schmidt}},
  \bibinfo {author} {\bibfnamefont {M.~S.}\ \bibnamefont {Gordon}},\ and\
  \bibinfo {author} {\bibfnamefont {K.}~\bibnamefont {Ruedenberg}},\ }\href
  {https://doi.org/10.1063/1.4840776} {\bibfield  {journal} {\bibinfo
  {journal} {J. Chem. Phys}\ }\textbf {\bibinfo {volume} {139}},\ \bibinfo
  {pages} {234107} (\bibinfo {year} {2013})}\BibitemShut {NoStop}%
\bibitem [{\citenamefont {Sayfutyarova}\ \emph {et~al.}(2017)\citenamefont
  {Sayfutyarova}, \citenamefont {Sun}, \citenamefont {Chan},\ and\
  \citenamefont {Knizia}}]{ElviraIAO}%
  \BibitemOpen
  \bibfield  {author} {\bibinfo {author} {\bibfnamefont {E.~R.}\ \bibnamefont
  {Sayfutyarova}}, \bibinfo {author} {\bibfnamefont {Q.}~\bibnamefont {Sun}},
  \bibinfo {author} {\bibfnamefont {G.~K.-L.}\ \bibnamefont {Chan}},\ and\
  \bibinfo {author} {\bibfnamefont {G.}~\bibnamefont {Knizia}},\ }\href
  {https://doi.org/10.1021/acs.jctc.7b00128} {\bibfield  {journal} {\bibinfo
  {journal} {J. Chem. Theory Comput}\ }\textbf {\bibinfo {volume} {13}},\
  \bibinfo {pages} {4063} (\bibinfo {year} {2017})}\BibitemShut {NoStop}%
\bibitem [{\citenamefont {Schneider}\ \emph {et~al.}(2016)\citenamefont
  {Schneider}, \citenamefont {Bistoni}, \citenamefont {Sparta}, \citenamefont
  {Saitow}, \citenamefont {Riplinger}, \citenamefont {Auer},\ and\
  \citenamefont {Neese}}]{SchneiderIAO}%
  \BibitemOpen
  \bibfield  {author} {\bibinfo {author} {\bibfnamefont {W.~B.}\ \bibnamefont
  {Schneider}}, \bibinfo {author} {\bibfnamefont {G.}~\bibnamefont {Bistoni}},
  \bibinfo {author} {\bibfnamefont {M.}~\bibnamefont {Sparta}}, \bibinfo
  {author} {\bibfnamefont {M.}~\bibnamefont {Saitow}}, \bibinfo {author}
  {\bibfnamefont {C.}~\bibnamefont {Riplinger}}, \bibinfo {author}
  {\bibfnamefont {A.~A.}\ \bibnamefont {Auer}},\ and\ \bibinfo {author}
  {\bibfnamefont {F.}~\bibnamefont {Neese}},\ }\href
  {https://doi.org/10.1021/acs.jctc.6b00523} {\bibfield  {journal} {\bibinfo
  {journal} {J. Chem. Theory Comput}\ }\textbf {\bibinfo {volume} {12}},\
  \bibinfo {pages} {4778} (\bibinfo {year} {2016})}\BibitemShut {NoStop}%
\bibitem [{\citenamefont {Barison}\ \emph {et~al.}(2022)\citenamefont
  {Barison}, \citenamefont {Galli},\ and\ \citenamefont
  {Motta}}]{barison2020quantum}%
  \BibitemOpen
  \bibfield  {author} {\bibinfo {author} {\bibfnamefont {S.}~\bibnamefont
  {Barison}}, \bibinfo {author} {\bibfnamefont {D.~E.}\ \bibnamefont {Galli}},\
  and\ \bibinfo {author} {\bibfnamefont {M.}~\bibnamefont {Motta}},\ }\href
  {https://journals.aps.org/pra/abstract/10.1103/PhysRevA.106.022404}
  {\bibfield  {journal} {\bibinfo  {journal} {Phys. Rev. A}\ }\textbf {\bibinfo
  {volume} {106}},\ \bibinfo {pages} {022404} (\bibinfo {year}
  {2022})}\BibitemShut {NoStop}%
\bibitem [{\citenamefont {Celotta}\ \emph {et~al.}(1974)\citenamefont
  {Celotta}, \citenamefont {Bennett},\ and\ \citenamefont
  {Hall}}]{celotta1974laser}%
  \BibitemOpen
  \bibfield  {author} {\bibinfo {author} {\bibfnamefont {R.}~\bibnamefont
  {Celotta}}, \bibinfo {author} {\bibfnamefont {R.}~\bibnamefont {Bennett}},\
  and\ \bibinfo {author} {\bibfnamefont {J.~L.}\ \bibnamefont {Hall}},\ }\href
  {https://aip.scitation.org/doi/10.1063/1.1681268} {\bibfield  {journal}
  {\bibinfo  {journal} {J. Chem. Phys}\ }\textbf {\bibinfo {volume} {60}},\
  \bibinfo {pages} {1740} (\bibinfo {year} {1974})}\BibitemShut {NoStop}%
\bibitem [{\citenamefont {Hotop}\ \emph {et~al.}(1974)\citenamefont {Hotop},
  \citenamefont {Patterson},\ and\ \citenamefont {Lineberger}}]{hotop1974high}%
  \BibitemOpen
  \bibfield  {author} {\bibinfo {author} {\bibfnamefont {H.}~\bibnamefont
  {Hotop}}, \bibinfo {author} {\bibfnamefont {T.}~\bibnamefont {Patterson}},\
  and\ \bibinfo {author} {\bibfnamefont {W.}~\bibnamefont {Lineberger}},\
  }\href {https://aip.scitation.org/doi/10.1063/1.1681279} {\bibfield
  {journal} {\bibinfo  {journal} {J. Chem. Phys}\ }\textbf {\bibinfo {volume}
  {60}},\ \bibinfo {pages} {1806} (\bibinfo {year} {1974})}\BibitemShut
  {NoStop}%
\bibitem [{\citenamefont {Schulz}\ \emph {et~al.}(1982)\citenamefont {Schulz},
  \citenamefont {Mead}, \citenamefont {Jones},\ and\ \citenamefont
  {Lineberger}}]{schulz1982oh}%
  \BibitemOpen
  \bibfield  {author} {\bibinfo {author} {\bibfnamefont {P.}~\bibnamefont
  {Schulz}}, \bibinfo {author} {\bibfnamefont {R.~D.}\ \bibnamefont {Mead}},
  \bibinfo {author} {\bibfnamefont {P.}~\bibnamefont {Jones}},\ and\ \bibinfo
  {author} {\bibfnamefont {W.}~\bibnamefont {Lineberger}},\ }\href
  {https://aip.scitation.org/doi/10.1063/1.443980} {\bibfield  {journal}
  {\bibinfo  {journal} {J. Chem. Phys}\ }\textbf {\bibinfo {volume} {77}},\
  \bibinfo {pages} {1153} (\bibinfo {year} {1982})}\BibitemShut {NoStop}%
\bibitem [{\citenamefont {Cade}(1967)}]{cade1967hartree}%
  \BibitemOpen
  \bibfield  {author} {\bibinfo {author} {\bibfnamefont {P.~E.}\ \bibnamefont
  {Cade}},\ }\href {https://aip.scitation.org/doi/10.1063/1.1703322} {\bibfield
   {journal} {\bibinfo  {journal} {J. Chem. Phys}\ }\textbf {\bibinfo {volume}
  {47}},\ \bibinfo {pages} {2390} (\bibinfo {year} {1967})}\BibitemShut
  {NoStop}%
\bibitem [{\citenamefont {Smith}\ \emph {et~al.}(1974)\citenamefont {Smith},
  \citenamefont {Chen},\ and\ \citenamefont {Simons}}]{smith1974theoretical}%
  \BibitemOpen
  \bibfield  {author} {\bibinfo {author} {\bibfnamefont {W.~D.}\ \bibnamefont
  {Smith}}, \bibinfo {author} {\bibfnamefont {T.-T.}\ \bibnamefont {Chen}},\
  and\ \bibinfo {author} {\bibfnamefont {J.}~\bibnamefont {Simons}},\ }\href
  {https://www.sciencedirect.com/science/article/abs/pii/0009261474802903}
  {\bibfield  {journal} {\bibinfo  {journal} {Chem. Phys. Lett}\ }\textbf
  {\bibinfo {volume} {27}},\ \bibinfo {pages} {499} (\bibinfo {year}
  {1974})}\BibitemShut {NoStop}%
\bibitem [{\citenamefont {Meyer}(1974)}]{meyer1974pno}%
  \BibitemOpen
  \bibfield  {author} {\bibinfo {author} {\bibfnamefont {W.}~\bibnamefont
  {Meyer}},\ }\href {https://link.springer.com/article/10.1007/BF00548478}
  {\bibfield  {journal} {\bibinfo  {journal} {Theor. Chim. Acta}\ }\textbf
  {\bibinfo {volume} {35}},\ \bibinfo {pages} {277} (\bibinfo {year}
  {1974})}\BibitemShut {NoStop}%
\bibitem [{\citenamefont {Sasaki}\ and\ \citenamefont
  {Yoshimine}(1974)}]{sasaki1974configuration}%
  \BibitemOpen
  \bibfield  {author} {\bibinfo {author} {\bibfnamefont {F.}~\bibnamefont
  {Sasaki}}\ and\ \bibinfo {author} {\bibfnamefont {M.}~\bibnamefont
  {Yoshimine}},\ }\href
  {https://journals.aps.org/pra/abstract/10.1103/PhysRevA.9.17} {\bibfield
  {journal} {\bibinfo  {journal} {Phys. Rev. A}\ }\textbf {\bibinfo {volume}
  {9}},\ \bibinfo {pages} {17} (\bibinfo {year} {1974})}\BibitemShut {NoStop}%
\bibitem [{\citenamefont {Rosmus}\ and\ \citenamefont
  {Meyer}(1978)}]{rosmus1978pno}%
  \BibitemOpen
  \bibfield  {author} {\bibinfo {author} {\bibfnamefont {P.}~\bibnamefont
  {Rosmus}}\ and\ \bibinfo {author} {\bibfnamefont {W.}~\bibnamefont {Meyer}},\
  }\href {https://aip.scitation.org/doi/abs/10.1063/1.436871} {\bibfield
  {journal} {\bibinfo  {journal} {J. Chem. Phys}\ }\textbf {\bibinfo {volume}
  {69}},\ \bibinfo {pages} {2745} (\bibinfo {year} {1978})}\BibitemShut
  {NoStop}%
\bibitem [{\citenamefont {Botch}\ and\ \citenamefont
  {Dunning~Jr}(1982)}]{botch1982theoretical}%
  \BibitemOpen
  \bibfield  {author} {\bibinfo {author} {\bibfnamefont {B.~H.}\ \bibnamefont
  {Botch}}\ and\ \bibinfo {author} {\bibfnamefont {T.~H.}\ \bibnamefont
  {Dunning~Jr}},\ }\href {https://aip.scitation.org/doi/10.1063/1.442959}
  {\bibfield  {journal} {\bibinfo  {journal} {J. Chem. Phys}\ }\textbf
  {\bibinfo {volume} {76}},\ \bibinfo {pages} {6046} (\bibinfo {year}
  {1982})}\BibitemShut {NoStop}%
\bibitem [{\citenamefont {Novoa}\ and\ \citenamefont
  {Mota}(1985)}]{novoa1985electron}%
  \BibitemOpen
  \bibfield  {author} {\bibinfo {author} {\bibfnamefont {J.~J.}\ \bibnamefont
  {Novoa}}\ and\ \bibinfo {author} {\bibfnamefont {F.}~\bibnamefont {Mota}},\
  }\href
  {https://www.sciencedirect.com/science/article/abs/pii/0009261485800476}
  {\bibfield  {journal} {\bibinfo  {journal} {Chem. Phys. Lett}\ }\textbf
  {\bibinfo {volume} {119}},\ \bibinfo {pages} {135} (\bibinfo {year}
  {1985})}\BibitemShut {NoStop}%
\bibitem [{\citenamefont {Raghavachari}(1985)}]{raghavachari1985basis}%
  \BibitemOpen
  \bibfield  {author} {\bibinfo {author} {\bibfnamefont {K.}~\bibnamefont
  {Raghavachari}},\ }\href {https://aip.scitation.org/doi/10.1063/1.448856}
  {\bibfield  {journal} {\bibinfo  {journal} {The Journal of chemical physics}\
  }\textbf {\bibinfo {volume} {82}},\ \bibinfo {pages} {4142} (\bibinfo {year}
  {1985})}\BibitemShut {NoStop}%
\bibitem [{\citenamefont {Baker}\ \emph {et~al.}(1986)\citenamefont {Baker},
  \citenamefont {Nobes},\ and\ \citenamefont {Radom}}]{baker1986evaluation}%
  \BibitemOpen
  \bibfield  {author} {\bibinfo {author} {\bibfnamefont {J.}~\bibnamefont
  {Baker}}, \bibinfo {author} {\bibfnamefont {R.~H.}\ \bibnamefont {Nobes}},\
  and\ \bibinfo {author} {\bibfnamefont {L.}~\bibnamefont {Radom}},\ }\href
  {https://onlinelibrary.wiley.com/doi/abs/10.1002/jcc.540070312} {\bibfield
  {journal} {\bibinfo  {journal} {J. Comput. Chem}\ }\textbf {\bibinfo {volume}
  {7}},\ \bibinfo {pages} {349} (\bibinfo {year} {1986})}\BibitemShut {NoStop}%
\bibitem [{\citenamefont {Chipman}(1986)}]{chipman1986electron}%
  \BibitemOpen
  \bibfield  {author} {\bibinfo {author} {\bibfnamefont {D.~M.}\ \bibnamefont
  {Chipman}},\ }\href {https://aip.scitation.org/doi/10.1063/1.450464}
  {\bibfield  {journal} {\bibinfo  {journal} {J. Chem. Phys}\ }\textbf
  {\bibinfo {volume} {84}},\ \bibinfo {pages} {1677} (\bibinfo {year}
  {1986})}\BibitemShut {NoStop}%
\bibitem [{\citenamefont {Dyall}(1995)}]{dyall1995choice}%
  \BibitemOpen
  \bibfield  {author} {\bibinfo {author} {\bibfnamefont {K.~G.}\ \bibnamefont
  {Dyall}},\ }\href {https://aip.scitation.org/doi/10.1063/1.469539} {\bibfield
   {journal} {\bibinfo  {journal} {J. Chem. Phys}\ }\textbf {\bibinfo {volume}
  {102}},\ \bibinfo {pages} {4909} (\bibinfo {year} {1995})}\BibitemShut
  {NoStop}%
\bibitem [{\citenamefont {Sun}\ \emph {et~al.}(2018)\citenamefont {Sun},
  \citenamefont {Berkelbach}, \citenamefont {Blunt}, \citenamefont {Booth},
  \citenamefont {Guo}, \citenamefont {Li}, \citenamefont {Liu}, \citenamefont
  {McClain}, \citenamefont {Sayfutyarova}, \citenamefont {Sharma} \emph
  {et~al.}}]{sun2018pyscf}%
  \BibitemOpen
  \bibfield  {author} {\bibinfo {author} {\bibfnamefont {Q.}~\bibnamefont
  {Sun}}, \bibinfo {author} {\bibfnamefont {T.~C.}\ \bibnamefont {Berkelbach}},
  \bibinfo {author} {\bibfnamefont {N.~S.}\ \bibnamefont {Blunt}}, \bibinfo
  {author} {\bibfnamefont {G.~H.}\ \bibnamefont {Booth}}, \bibinfo {author}
  {\bibfnamefont {S.}~\bibnamefont {Guo}}, \bibinfo {author} {\bibfnamefont
  {Z.}~\bibnamefont {Li}}, \bibinfo {author} {\bibfnamefont {J.}~\bibnamefont
  {Liu}}, \bibinfo {author} {\bibfnamefont {J.~D.}\ \bibnamefont {McClain}},
  \bibinfo {author} {\bibfnamefont {E.~R.}\ \bibnamefont {Sayfutyarova}},
  \bibinfo {author} {\bibfnamefont {S.}~\bibnamefont {Sharma}}, \emph
  {et~al.},\ }\href {https://onlinelibrary.wiley.com/doi/abs/10.1002/wcms.1340}
  {\bibfield  {journal} {\bibinfo  {journal} {WIREs Comput. Mol. Sci}\ }\textbf
  {\bibinfo {volume} {8}},\ \bibinfo {pages} {e1340} (\bibinfo {year}
  {2018})}\BibitemShut {NoStop}%
\bibitem [{\citenamefont {Sun}\ \emph {et~al.}(2020)\citenamefont {Sun} \emph
  {et~al.}}]{sun2020recent}%
  \BibitemOpen
  \bibfield  {author} {\bibinfo {author} {\bibfnamefont {Q.}~\bibnamefont
  {Sun}} \emph {et~al.},\ }\href {https://doi.org/10.1063/5.0006074} {\bibfield
   {journal} {\bibinfo  {journal} {J. Chem. Phys}\ }\textbf {\bibinfo {volume}
  {153}},\ \bibinfo {pages} {024109} (\bibinfo {year} {2020})}\BibitemShut
  {NoStop}%
\bibitem [{\citenamefont {Aleksandrowicz}\ \emph {et~al.}(2019)\citenamefont
  {Aleksandrowicz}, \citenamefont {Alexander}, \citenamefont {Barkoutsos},
  \citenamefont {Bello}, \citenamefont {Ben-Haim}, \citenamefont {Bucher},
  \citenamefont {Cabrera-Hern{\'a}ndez}, \citenamefont {Carballo-Franquis},
  \citenamefont {Chen}, \citenamefont {Chen} \emph
  {et~al.}}]{aleksandrowicz2019qiskit}%
  \BibitemOpen
  \bibfield  {author} {\bibinfo {author} {\bibfnamefont {G.}~\bibnamefont
  {Aleksandrowicz}}, \bibinfo {author} {\bibfnamefont {T.}~\bibnamefont
  {Alexander}}, \bibinfo {author} {\bibfnamefont {P.}~\bibnamefont
  {Barkoutsos}}, \bibinfo {author} {\bibfnamefont {L.}~\bibnamefont {Bello}},
  \bibinfo {author} {\bibfnamefont {Y.}~\bibnamefont {Ben-Haim}}, \bibinfo
  {author} {\bibfnamefont {D.}~\bibnamefont {Bucher}}, \bibinfo {author}
  {\bibfnamefont {F.}~\bibnamefont {Cabrera-Hern{\'a}ndez}}, \bibinfo {author}
  {\bibfnamefont {J.}~\bibnamefont {Carballo-Franquis}}, \bibinfo {author}
  {\bibfnamefont {A.}~\bibnamefont {Chen}}, \bibinfo {author} {\bibfnamefont
  {C.}~\bibnamefont {Chen}}, \emph {et~al.},\ }\href
  {https://zenodo.org/record/2562111#.XhA8qi2ZPyI} {\bibfield  {journal}
  {\bibinfo  {journal} {Zenodo}\ }\textbf {\bibinfo {volume} {16}} (\bibinfo
  {year} {2019})}\BibitemShut {NoStop}%
\bibitem [{\citenamefont {Bravyi}\ \emph {et~al.}(2017)\citenamefont {Bravyi},
  \citenamefont {Gambetta}, \citenamefont {Mezzacapo},\ and\ \citenamefont
  {Temme}}]{bravyi2017tapering}%
  \BibitemOpen
  \bibfield  {author} {\bibinfo {author} {\bibfnamefont {S.}~\bibnamefont
  {Bravyi}}, \bibinfo {author} {\bibfnamefont {J.~M.}\ \bibnamefont
  {Gambetta}}, \bibinfo {author} {\bibfnamefont {A.}~\bibnamefont
  {Mezzacapo}},\ and\ \bibinfo {author} {\bibfnamefont {K.}~\bibnamefont
  {Temme}},\ }\href {https://arxiv.org/abs/1701.08213} {\bibfield  {journal}
  {\bibinfo  {journal} {arXiv:1701.08213}\ } (\bibinfo {year}
  {2017})}\BibitemShut {NoStop}%
\bibitem [{\citenamefont {Setia}\ \emph {et~al.}(2020)\citenamefont {Setia},
  \citenamefont {Chen}, \citenamefont {Rice}, \citenamefont {Mezzacapo},
  \citenamefont {Pistoia},\ and\ \citenamefont
  {Whitfield}}]{setia2019reducing}%
  \BibitemOpen
  \bibfield  {author} {\bibinfo {author} {\bibfnamefont {K.}~\bibnamefont
  {Setia}}, \bibinfo {author} {\bibfnamefont {R.}~\bibnamefont {Chen}},
  \bibinfo {author} {\bibfnamefont {J.~E.}\ \bibnamefont {Rice}}, \bibinfo
  {author} {\bibfnamefont {A.}~\bibnamefont {Mezzacapo}}, \bibinfo {author}
  {\bibfnamefont {M.}~\bibnamefont {Pistoia}},\ and\ \bibinfo {author}
  {\bibfnamefont {J.~D.}\ \bibnamefont {Whitfield}},\ }\href
  {https://pubs.acs.org/doi/10.1021/acs.jctc.0c00113} {\bibfield  {journal}
  {\bibinfo  {journal} {J. Chem. Theory Comput}\ }\textbf {\bibinfo {volume}
  {16}},\ \bibinfo {pages} {6091} (\bibinfo {year} {2020})}\BibitemShut
  {NoStop}%
\bibitem [{\citenamefont {Barkoutsos}\ \emph {et~al.}(2018)\citenamefont
  {Barkoutsos}, \citenamefont {Gonthier}, \citenamefont {Sokolov},
  \citenamefont {Moll}, \citenamefont {Salis}, \citenamefont {Fuhrer},
  \citenamefont {Ganzhorn}, \citenamefont {Egger}, \citenamefont {Troyer},
  \citenamefont {Mezzacapo}, \citenamefont {Filipp},\ and\ \citenamefont
  {Tavernelli}}]{barkoutsos2018quantum}%
  \BibitemOpen
  \bibfield  {author} {\bibinfo {author} {\bibfnamefont {P.~K.}\ \bibnamefont
  {Barkoutsos}}, \bibinfo {author} {\bibfnamefont {J.~F.}\ \bibnamefont
  {Gonthier}}, \bibinfo {author} {\bibfnamefont {I.}~\bibnamefont {Sokolov}},
  \bibinfo {author} {\bibfnamefont {N.}~\bibnamefont {Moll}}, \bibinfo {author}
  {\bibfnamefont {G.}~\bibnamefont {Salis}}, \bibinfo {author} {\bibfnamefont
  {A.}~\bibnamefont {Fuhrer}}, \bibinfo {author} {\bibfnamefont
  {M.}~\bibnamefont {Ganzhorn}}, \bibinfo {author} {\bibfnamefont {D.~J.}\
  \bibnamefont {Egger}}, \bibinfo {author} {\bibfnamefont {M.}~\bibnamefont
  {Troyer}}, \bibinfo {author} {\bibfnamefont {A.}~\bibnamefont {Mezzacapo}},
  \bibinfo {author} {\bibfnamefont {S.}~\bibnamefont {Filipp}},\ and\ \bibinfo
  {author} {\bibfnamefont {I.}~\bibnamefont {Tavernelli}},\ }\href
  {https://doi.org/10.1103/PhysRevA.98.022322} {\bibfield  {journal} {\bibinfo
  {journal} {Phys. Rev. A}\ }\textbf {\bibinfo {volume} {98}},\ \bibinfo
  {pages} {022322} (\bibinfo {year} {2018})}\BibitemShut {NoStop}%
\bibitem [{\citenamefont {Zhu}\ \emph {et~al.}(1997)\citenamefont {Zhu},
  \citenamefont {Byrd}, \citenamefont {Lu},\ and\ \citenamefont
  {Nocedal}}]{zhu1997algorithm}%
  \BibitemOpen
  \bibfield  {author} {\bibinfo {author} {\bibfnamefont {C.}~\bibnamefont
  {Zhu}}, \bibinfo {author} {\bibfnamefont {R.~H.}\ \bibnamefont {Byrd}},
  \bibinfo {author} {\bibfnamefont {P.}~\bibnamefont {Lu}},\ and\ \bibinfo
  {author} {\bibfnamefont {J.}~\bibnamefont {Nocedal}},\ }\href
  {https://doi.org/10.1145/279232.279236} {\bibfield  {journal} {\bibinfo
  {journal} {ACM Trans. Math. Softw.}\ }\textbf {\bibinfo {volume} {23}},\
  \bibinfo {pages} {550–560} (\bibinfo {year} {1997})}\BibitemShut {NoStop}%
\bibitem [{\citenamefont {Morales}\ and\ \citenamefont
  {Nocedal}(2011)}]{morales2011remark}%
  \BibitemOpen
  \bibfield  {author} {\bibinfo {author} {\bibfnamefont {J.~L.}\ \bibnamefont
  {Morales}}\ and\ \bibinfo {author} {\bibfnamefont {J.}~\bibnamefont
  {Nocedal}},\ }\href {https://doi.org/10.1145/2049662.2049669} {\bibfield
  {journal} {\bibinfo  {journal} {ACM Trans. Math. Softw.}\ }\textbf {\bibinfo
  {volume} {38}},\ \bibinfo {pages} {7} (\bibinfo {year} {2011})}\BibitemShut
  {NoStop}%
\bibitem [{\citenamefont {Temme}\ \emph {et~al.}(2017)\citenamefont {Temme},
  \citenamefont {Bravyi},\ and\ \citenamefont {Gambetta}}]{temme2017error}%
  \BibitemOpen
  \bibfield  {author} {\bibinfo {author} {\bibfnamefont {K.}~\bibnamefont
  {Temme}}, \bibinfo {author} {\bibfnamefont {S.}~\bibnamefont {Bravyi}},\ and\
  \bibinfo {author} {\bibfnamefont {J.~M.}\ \bibnamefont {Gambetta}},\ }\href
  {https://journals.aps.org/prl/abstract/10.1103/PhysRevLett.119.180509}
  {\bibfield  {journal} {\bibinfo  {journal} {Phys. Rev. Lett}\ }\textbf
  {\bibinfo {volume} {119}},\ \bibinfo {pages} {180509} (\bibinfo {year}
  {2017})}\BibitemShut {NoStop}%
\bibitem [{\citenamefont {Kandala}\ \emph {et~al.}(2019)\citenamefont
  {Kandala}, \citenamefont {Temme}, \citenamefont {C{\'o}rcoles}, \citenamefont
  {Mezzacapo}, \citenamefont {Chow},\ and\ \citenamefont
  {Gambetta}}]{kandala2019error}%
  \BibitemOpen
  \bibfield  {author} {\bibinfo {author} {\bibfnamefont {A.}~\bibnamefont
  {Kandala}}, \bibinfo {author} {\bibfnamefont {K.}~\bibnamefont {Temme}},
  \bibinfo {author} {\bibfnamefont {A.~D.}\ \bibnamefont {C{\'o}rcoles}},
  \bibinfo {author} {\bibfnamefont {A.}~\bibnamefont {Mezzacapo}}, \bibinfo
  {author} {\bibfnamefont {J.~M.}\ \bibnamefont {Chow}},\ and\ \bibinfo
  {author} {\bibfnamefont {J.~M.}\ \bibnamefont {Gambetta}},\ }\href
  {https://www.nature.com/articles/s41586-019-1040-7?platform=hootsuite}
  {\bibfield  {journal} {\bibinfo  {journal} {Nature}\ }\textbf {\bibinfo
  {volume} {567}},\ \bibinfo {pages} {491} (\bibinfo {year}
  {2019})}\BibitemShut {NoStop}%
\bibitem [{\citenamefont {Bravyi}\ \emph {et~al.}(2021)\citenamefont {Bravyi},
  \citenamefont {Sheldon}, \citenamefont {Kandala}, \citenamefont {Mckay},\
  and\ \citenamefont {Gambetta}}]{bravyi2020mitigating}%
  \BibitemOpen
  \bibfield  {author} {\bibinfo {author} {\bibfnamefont {S.}~\bibnamefont
  {Bravyi}}, \bibinfo {author} {\bibfnamefont {S.}~\bibnamefont {Sheldon}},
  \bibinfo {author} {\bibfnamefont {A.}~\bibnamefont {Kandala}}, \bibinfo
  {author} {\bibfnamefont {D.~C.}\ \bibnamefont {Mckay}},\ and\ \bibinfo
  {author} {\bibfnamefont {J.~M.}\ \bibnamefont {Gambetta}},\ }\href
  {https://journals.aps.org/pra/abstract/10.1103/PhysRevA.103.042605}
  {\bibfield  {journal} {\bibinfo  {journal} {Phys. Rev. A}\ }\textbf {\bibinfo
  {volume} {103}},\ \bibinfo {pages} {042605} (\bibinfo {year}
  {2021})}\BibitemShut {NoStop}%
\bibitem [{\citenamefont {Nation}\ \emph {et~al.}(2021)\citenamefont {Nation},
  \citenamefont {Kang}, \citenamefont {Sundaresan},\ and\ \citenamefont
  {Gambetta}}]{nation2021scalable}%
  \BibitemOpen
  \bibfield  {author} {\bibinfo {author} {\bibfnamefont {P.~D.}\ \bibnamefont
  {Nation}}, \bibinfo {author} {\bibfnamefont {H.}~\bibnamefont {Kang}},
  \bibinfo {author} {\bibfnamefont {N.}~\bibnamefont {Sundaresan}},\ and\
  \bibinfo {author} {\bibfnamefont {J.~M.}\ \bibnamefont {Gambetta}},\ }\href
  {https://journals.aps.org/prxquantum/abstract/10.1103/PRXQuantum.2.040326}
  {\bibfield  {journal} {\bibinfo  {journal} {PRX Quantum}\ }\textbf {\bibinfo
  {volume} {2}},\ \bibinfo {pages} {040326} (\bibinfo {year}
  {2021})}\BibitemShut {NoStop}%
\bibitem [{\citenamefont {Dumitrescu}\ \emph {et~al.}(2018)\citenamefont
  {Dumitrescu}, \citenamefont {McCaskey}, \citenamefont {Hagen}, \citenamefont
  {Jansen}, \citenamefont {Morris}, \citenamefont {Papenbrock}, \citenamefont
  {Pooser}, \citenamefont {Dean},\ and\ \citenamefont
  {Lougovski}}]{dumitrescu2018cloud}%
  \BibitemOpen
  \bibfield  {author} {\bibinfo {author} {\bibfnamefont {E.~F.}\ \bibnamefont
  {Dumitrescu}}, \bibinfo {author} {\bibfnamefont {A.~J.}\ \bibnamefont
  {McCaskey}}, \bibinfo {author} {\bibfnamefont {G.}~\bibnamefont {Hagen}},
  \bibinfo {author} {\bibfnamefont {G.~R.}\ \bibnamefont {Jansen}}, \bibinfo
  {author} {\bibfnamefont {T.~D.}\ \bibnamefont {Morris}}, \bibinfo {author}
  {\bibfnamefont {T.}~\bibnamefont {Papenbrock}}, \bibinfo {author}
  {\bibfnamefont {R.~C.}\ \bibnamefont {Pooser}}, \bibinfo {author}
  {\bibfnamefont {D.~J.}\ \bibnamefont {Dean}},\ and\ \bibinfo {author}
  {\bibfnamefont {P.}~\bibnamefont {Lougovski}},\ }\href
  {https://doi.org/10.1103/PhysRevLett.120.210501} {\bibfield  {journal}
  {\bibinfo  {journal} {Phys. Rev. Lett.}\ }\textbf {\bibinfo {volume} {120}},\
  \bibinfo {pages} {210501} (\bibinfo {year} {2018})}\BibitemShut {NoStop}%
\bibitem [{\citenamefont {Stamatopoulos}\ \emph {et~al.}(2020)\citenamefont
  {Stamatopoulos}, \citenamefont {Egger}, \citenamefont {Sun}, \citenamefont
  {Zoufal}, \citenamefont {Iten}, \citenamefont {Shen},\ and\ \citenamefont
  {Woerner}}]{stamatopoulos2019option}%
  \BibitemOpen
  \bibfield  {author} {\bibinfo {author} {\bibfnamefont {N.}~\bibnamefont
  {Stamatopoulos}}, \bibinfo {author} {\bibfnamefont {D.~J.}\ \bibnamefont
  {Egger}}, \bibinfo {author} {\bibfnamefont {Y.}~\bibnamefont {Sun}}, \bibinfo
  {author} {\bibfnamefont {C.}~\bibnamefont {Zoufal}}, \bibinfo {author}
  {\bibfnamefont {R.}~\bibnamefont {Iten}}, \bibinfo {author} {\bibfnamefont
  {N.}~\bibnamefont {Shen}},\ and\ \bibinfo {author} {\bibfnamefont
  {S.}~\bibnamefont {Woerner}},\ }\href
  {https://dx.doi.org/10.22331/q-2020-07-06-291} {\bibfield  {journal}
  {\bibinfo  {journal} {Quantum}\ }\textbf {\bibinfo {volume} {4}},\ \bibinfo
  {pages} {291} (\bibinfo {year} {2020})}\BibitemShut {NoStop}%
\bibitem [{\citenamefont {Hehre}\ \emph {et~al.}(1972)\citenamefont {Hehre},
  \citenamefont {Ditchfield},\ and\ \citenamefont {Pople}}]{hehre1972self}%
  \BibitemOpen
  \bibfield  {author} {\bibinfo {author} {\bibfnamefont {W.~J.}\ \bibnamefont
  {Hehre}}, \bibinfo {author} {\bibfnamefont {R.}~\bibnamefont {Ditchfield}},\
  and\ \bibinfo {author} {\bibfnamefont {J.~A.}\ \bibnamefont {Pople}},\ }\href
  {https://aip.scitation.org/doi/10.1063/1.1677527} {\bibfield  {journal}
  {\bibinfo  {journal} {J. Chem. Phys}\ }\textbf {\bibinfo {volume} {56}},\
  \bibinfo {pages} {2257} (\bibinfo {year} {1972})}\BibitemShut {NoStop}%
\bibitem [{\citenamefont {Dunning~Jr}(1989)}]{dunning1989gaussian}%
  \BibitemOpen
  \bibfield  {author} {\bibinfo {author} {\bibfnamefont {T.~H.}\ \bibnamefont
  {Dunning~Jr}},\ }\href {https://aip.scitation.org/doi/10.1063/1.456153}
  {\bibfield  {journal} {\bibinfo  {journal} {J. Chem. Phys}\ }\textbf
  {\bibinfo {volume} {90}},\ \bibinfo {pages} {1007} (\bibinfo {year}
  {1989})}\BibitemShut {NoStop}%
\bibitem [{\citenamefont {Johnson~III}(2019)}]{johnson1999nist}%
  \BibitemOpen
  \bibfield  {author} {\bibinfo {author} {\bibfnamefont {R.~D.}\ \bibnamefont
  {Johnson~III}},\ }\href {http://cccbdb.nist.gov/} {\emph {\bibinfo {title}
  {NIST 101. Computational chemistry comparison and benchmark database}}},\
  \bibinfo {type} {Tech. Rep.}\ (\bibinfo  {institution} {National Institute of
  Standards and Technology},\ \bibinfo {year} {2019})\BibitemShut {NoStop}%
\bibitem [{\citenamefont {Feller}(1992)}]{feller1992application}%
  \BibitemOpen
  \bibfield  {author} {\bibinfo {author} {\bibfnamefont {D.}~\bibnamefont
  {Feller}},\ }\href {https://aip.scitation.org/doi/10.1063/1.462652}
  {\bibfield  {journal} {\bibinfo  {journal} {J. Chem. Phys}\ }\textbf
  {\bibinfo {volume} {96}},\ \bibinfo {pages} {6104} (\bibinfo {year}
  {1992})}\BibitemShut {NoStop}%
\bibitem [{\citenamefont {Helgaker}\ \emph {et~al.}(1997)\citenamefont
  {Helgaker}, \citenamefont {Klopper}, \citenamefont {Koch},\ and\
  \citenamefont {Noga}}]{helgaker1997basis}%
  \BibitemOpen
  \bibfield  {author} {\bibinfo {author} {\bibfnamefont {T.}~\bibnamefont
  {Helgaker}}, \bibinfo {author} {\bibfnamefont {W.}~\bibnamefont {Klopper}},
  \bibinfo {author} {\bibfnamefont {H.}~\bibnamefont {Koch}},\ and\ \bibinfo
  {author} {\bibfnamefont {J.}~\bibnamefont {Noga}},\ }\href
  {https://aip.scitation.org/doi/10.1063/1.473863} {\bibfield  {journal}
  {\bibinfo  {journal} {J. Chem. Phys}\ }\textbf {\bibinfo {volume} {106}},\
  \bibinfo {pages} {9639} (\bibinfo {year} {1997})}\BibitemShut {NoStop}%
\bibitem [{\citenamefont {Andersson}\ \emph {et~al.}(1990)\citenamefont
  {Andersson}, \citenamefont {Malmqvist}, \citenamefont {Roos}, \citenamefont
  {Sadlej},\ and\ \citenamefont {Wolinski}}]{andersson1990second}%
  \BibitemOpen
  \bibfield  {author} {\bibinfo {author} {\bibfnamefont {K.}~\bibnamefont
  {Andersson}}, \bibinfo {author} {\bibfnamefont {P.~A.}\ \bibnamefont
  {Malmqvist}}, \bibinfo {author} {\bibfnamefont {B.~O.}\ \bibnamefont {Roos}},
  \bibinfo {author} {\bibfnamefont {A.~J.}\ \bibnamefont {Sadlej}},\ and\
  \bibinfo {author} {\bibfnamefont {K.}~\bibnamefont {Wolinski}},\ }\href
  {https://pubs.acs.org/doi/10.1021/j100377a012} {\bibfield  {journal}
  {\bibinfo  {journal} {J. Phys. Chem}\ }\textbf {\bibinfo {volume} {94}},\
  \bibinfo {pages} {5483} (\bibinfo {year} {1990})}\BibitemShut {NoStop}%
\bibitem [{\citenamefont {Andersson}\ \emph {et~al.}(1992)\citenamefont
  {Andersson}, \citenamefont {Malmqvist},\ and\ \citenamefont
  {Roos}}]{andersson1992second}%
  \BibitemOpen
  \bibfield  {author} {\bibinfo {author} {\bibfnamefont {K.}~\bibnamefont
  {Andersson}}, \bibinfo {author} {\bibfnamefont {P.-{\AA}.}\ \bibnamefont
  {Malmqvist}},\ and\ \bibinfo {author} {\bibfnamefont {B.~O.}\ \bibnamefont
  {Roos}},\ }\href
  {https://aip.scitation.org/doi/abs/10.1063/1.462209?journalCode=jcp}
  {\bibfield  {journal} {\bibinfo  {journal} {J. Chem. Phys}\ }\textbf
  {\bibinfo {volume} {96}},\ \bibinfo {pages} {1218} (\bibinfo {year}
  {1992})}\BibitemShut {NoStop}%
\end{thebibliography}

%

\end{document}